\documentclass[conference]{IEEEtran}
\IEEEoverridecommandlockouts
\usepackage{cite}
\usepackage{amsmath,amssymb,amsfonts}
\usepackage{algorithmic}
\usepackage{graphicx}
\usepackage{textcomp}
\usepackage{xcolor}
\def\BibTeX{{\rm B\kern-.05em{\sc i\kern-.025em b}\kern-.08em
    T\kern-.1667em\lower.7ex\hbox{E}\kern-.125emX}}
\begin{document}

%\title{Quantum Transform Image Classification, and Generating and Detecting Quantum Product States Using Quantum GANs\\
%}
\title{A quantum GAN for entanglement detection and image classification}

\author{\IEEEauthorblockN{1\textsuperscript{st} James E Steck }
\IEEEauthorblockA{\textit{Aerospace Engineering} \\
\textit{Wichita State University}\\
Wichita, KS USA \\
james.steck@wichita.edu}
\and
\IEEEauthorblockN{2\textsuperscript{nd} Elizabeth C Behrman}
\IEEEauthorblockA{\textit{Mathematics, Statistics and Physics} \\
\textit{Wichita State University}\\
Wichita, KS USA \\
ecbehrman@gmail.com}
\and
\IEEEauthorblockN{3\textsuperscript{rd} Nathan L Thompson}
\IEEEauthorblockA{\textit{Mathematics, Statistics and Physics} \\
\textit{Wichita State University}\\
Wichita, KS USA \\
nathan.thompson@wichita.edu}

}

\maketitle

\begin{abstract}
 We build on our prior quantum machine learning work to design and train a quantum GAN. We first show that it is a truly quantum network by showing that it can successfully classify the entanglement of input quantum states, an open NP-hard problem that is purely quantum mechanical and has no classical analog. We then apply the method of our quantum discriminator in our QGAN to image classification, and demonstrate both efficient encoding of images into quantum states and successful discrimination of database images. 

\end{abstract}


\section{Introduction}
For over thirty years now the scientific community has known that quantum computing has the potential to take a giant leap forward in computing capabilities, solving problems difficult or even impossible to solve classically. Recently the broader community’s interest has also been engaged, and startups and investment are growing. But there remain major obstacles to the implementation of macroscopic quantum computing: hardware problems of noise, decoherence, and scaling; and software problems of error correction and, most important, algorithm construction. Finding truly quantum algorithms turns out to be quite difficult. There are still only a very few. Most are based on one of three methods: quantum Fourier transform methods, like Shor’s \cite{origb1}; amplitude amplification, like Grover’s \cite{origb2}; and quantum walks \cite{origb3}. Shor’s, and some quantum walk algorithms, provide an exponential advantage over the best-known classical algorithm, and this is usually what people are thinking of for the really hard problems. But amplitude amplification only gives a quadratic speedup. It is not known whether a quantum advantage even exists for broad classes of problems \cite{origb4} \cite{origb5}, or, if it does, whether it will be exponential or somewhat weaker.  Even when the existence of an advantage can be proved, explicit construction of the algorithm is still not easy let alone obvious.

Now, the usual algorithmic approach is a kind of “building block” strategy, in which the procedure is formulated as a sequence of steps (quantum gates) from a universal set, e.g., a sequence of CNOT, Hadamard, and phase shift gates. This is of course exactly analogous to the way in which we usually do classical computing, as a series of logical gates operating on bits. But there is another computing paradigm we could follow: distributed computing, the approach of biological and of artificial neural networks. Since the 1990s our research group has been investigating this different approach, a combining of quantum computing and artificial neural networks, as an alternative to the building block paradigm. With this approach, the quantum systems itself learns how to solve the problem, designing its own algorithm in a sense. Moreover, we have shown \cite{origb6} \cite{origb7} that not only does this eliminate the program design obstacle, but also gives us near-automatic scaling \cite{origb8}, robustness to noise and to decoherence \cite{origb9}, and speedup over classical learning \cite{origb10}. 

The basic idea is that a quantum system can itself act as a neural network: The state of the system at the initial time is the “input”; a measurement (observable) on the system at the final time is the “output”; the states of the system at intermediate times are the hidden layers of the network. If we know enough about the computation desired to be able to construct a comprehensive set of input-output pairs from which the net can generalize, then, we can use techniques of machine learning to bypass the algorithm-construction problem. 

\section{Generative Adversarial Networks - GANs}

\subsection{Classical GANs}
\begin{figure}[htbp]
\centerline{\includegraphics{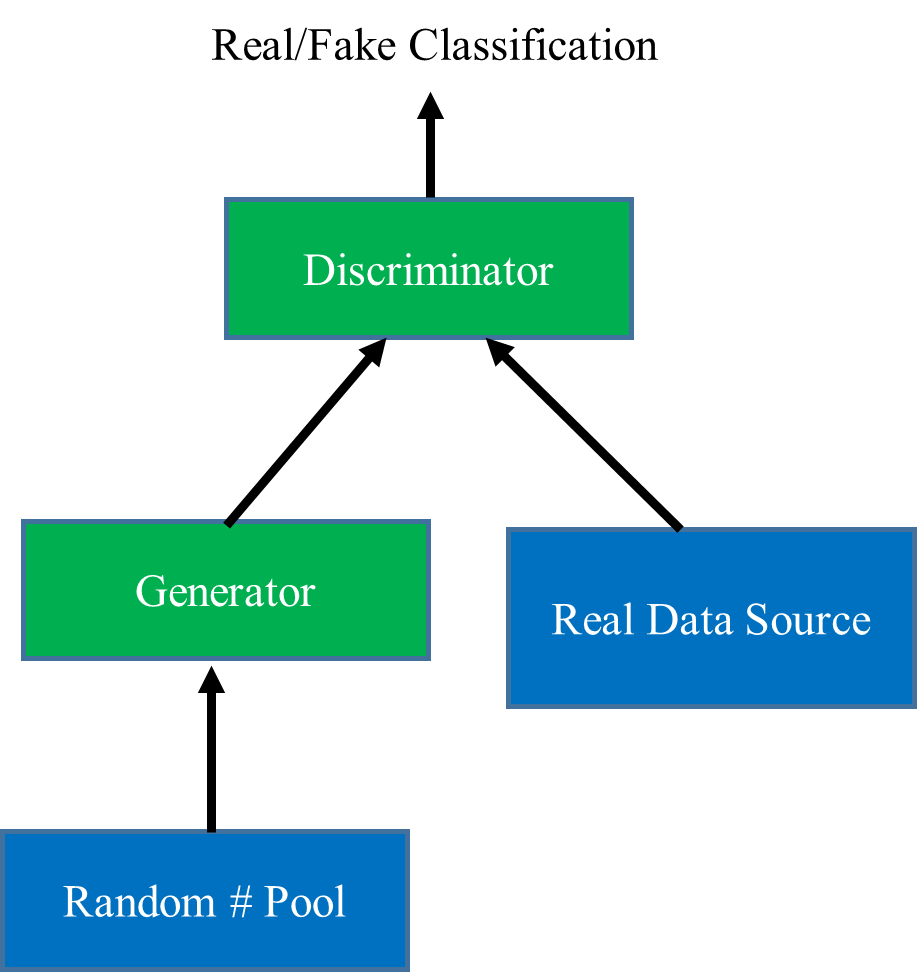}}
\caption{GAN Block Diagram.}
\label{Classical-GAN}
\end{figure}

Machine learning, via shallow and then deep convolution neural networks, has been used for a huge variety of AI tasks since the late 80’s, including data processing, recognition, classification, and novel data generation \cite{NeuralNet_Textbook}. In 2014 Ian J. Goodfellow et al. \cite{origb28} introduced the Generative Adversarial Network, or GAN. This is a structure containing two deep networks represented by the green blocks in Fig. \ref{Classical-GAN}: a Generator network that inputs a vector from a random pool and learns to generate fake data, and a Discriminator network that learns to classify data as either a generated fake or belonging to a real data source.  The learning occurs by pitting the Discriminator and the Generator against each other in a min/max zero-sum game: the Discriminator learning cost function is its error in classifying the fake and real data, and the Generator cost function is how correctly the Discriminator classified the fake data it generated.  GANs have been successful in image synthesis, translation, enhancement, speech synthesis, composition and sparse data set augmentation [29, 30]. Others have used GANs to create artwork in the style of famous artists, some of which sold at auction for six figures.  One of the most visually impressive and successful GANs are the image generation work of Karras et al. in 2018 at Nvidia [31, 32].  This motivated us toward applying our prior work on machine learning for quantum computing to create a quantum GAN or QGAN that mimics the approach of Karras’s stylegan [33] but takes advantage of a quantum computing framework and its ability to represent high resolution images in a relatively few number of quantum bits (qubits), compared to the large number of classical bits required.

\subsection{Quantum GANs}

\begin{figure}[htbp]
\centerline{\includegraphics{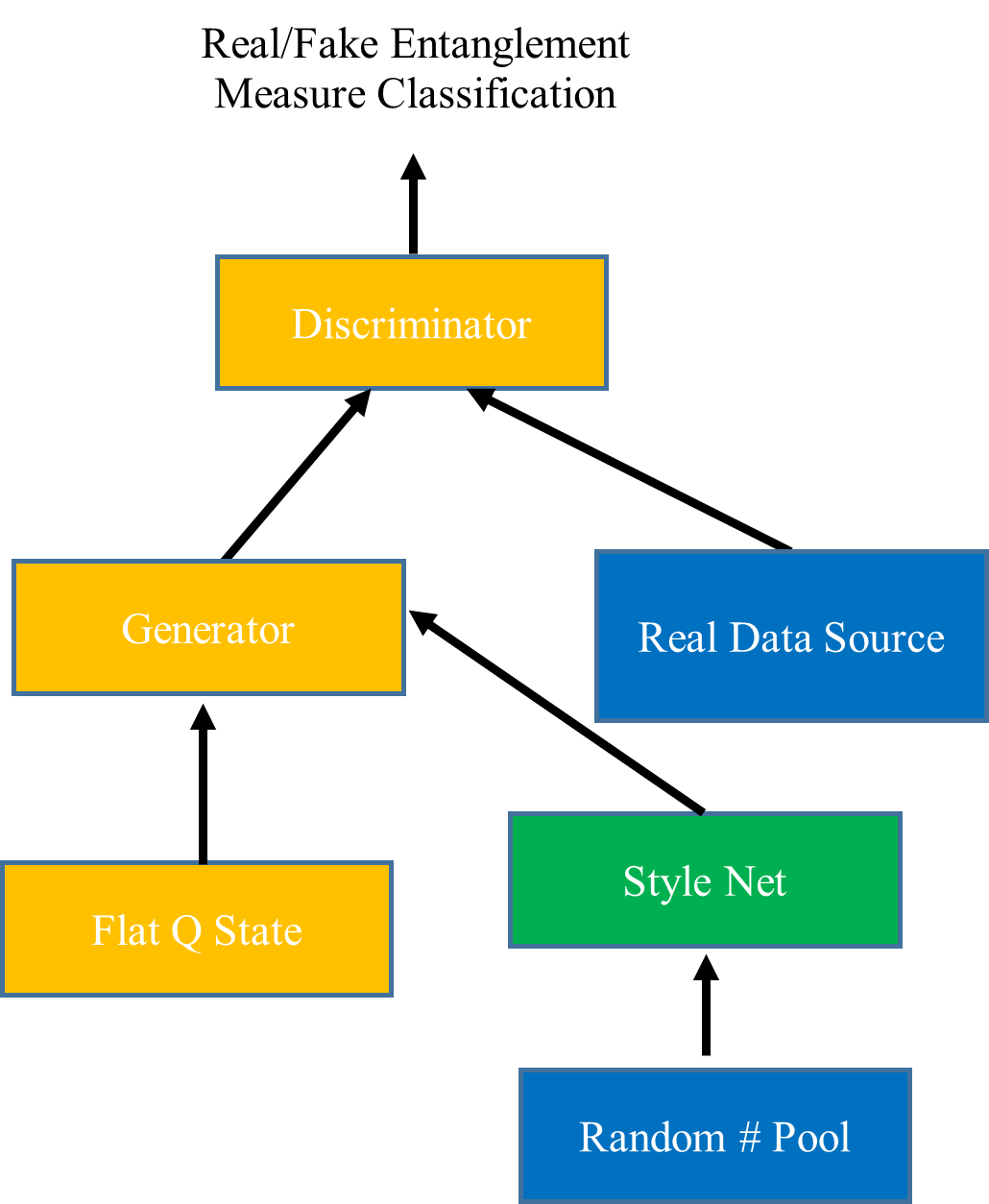}}
\caption{Quantum GAN Block Diagram.}
\label{Quantum-GAN}
\end{figure}

Shortly after Karras’s success in image generation with GANs, Seth Lloyd at MIT \cite{origb6} proposed the theory of a Quantum GAN and showed that, because of the nature of quantum systems, it is much easier to prove the convergence properties of a QGAN. A quantum GAN would replace the deep convolutional classical network of classical GANS needing a massive NVIDIA server cluster implementing the deep network with a quantum computer of much lesser complexity.  Pierro-Luc Dallaire-Demers at Xanadu \cite{origb24} published a QGAN structure based on the quantum computational gate model used in IBM’s as well as other Noisy Intermediate Scale Quantum (NISQ) hardware. This gate model is how most algorithms are created for quantum computation, yet, with this model, it is difficult in practice to program a quantum computer for more than just a few specific tasks.  Over the past 20 years, we, at Wichita State University, have developed machine learning methods for quantum systems that circumvents the gate models [25-31], and propose creating a QGAN, Fig. \ref{Quantum-GAN}, with quantum systems as Generator and Discriminator (orange blocks) that are trained with our methods and cost functions that use quantum entanglement measures.

As a demonstration of the purely quantum processing ability of the QGAN, we attack an important outstanding NP-hard problem: that of determining  “separability”, which is a uniquely quantum problem. All classical states are separable: that is, if a system consists of subsystems A and B, their joint state is the product of the state of A with the state of B. Quantum systems, on the other hand, can be NON-separable, or what is called “entangled”: their joint state can be such that neither subsystem A nor subsystem B has an individual state. We will use the real/fake Generator/Discriminator competition to find the boundary of the subspace of product states, identifying “real” with “product.” This will demonstrate that a QGAN does a purely quantum task, and, if successful, solve an important problem in the field of quantum information, and demonstrate the data creation and classification ability of the QGAN. 

A crucial feature required to perform quantum image processing on a quantum computer is transforming an M×M real image to a quantum state. The obvious method is to map each pixel of the image onto the amplitude of an N-qubit quantum state with N = $\log_2(M \times M)$. While very brute force, this method still significantly compresses the size of the image representation. Other approaches map an image onto a 2D array of rotation gates in an ansatz thereby compressing the image into an M qubit state, which is an efficient step for image classification using a quantum gate model computer, yet loses much of the image pixel distribution information.  An efficient approach that maintains all the pixel information and is a reversible transform is to take a 2D FFT of the image, which results in an MxM matrix of complex values that can be rearranged into a Hermitian quantum density matrix of a quantum state of N qubits with N = log2(M), an even more significant quantum compression of the image representation. A 1024 by 1024 image can thus be represented on and processed by a quantum computer with only 10 qubits. Once that is done, a quantum computer can be trained to perform image processing and feature recognition on the image via the machine learning methods that we have developed for quantum systems. Thus, the QGAN architecture we propose here could be applied to a large resource and computation speedup for many kinds of image creation and processing tasks, such as satellite surveillance, medical imaging, facial recognition, and so on.

In previous work we devised a quantum machine learning method to learn a time-dependent Hamiltonian for a multiqubit system such that a chosen measurement at the final time is a witness of the entanglement of the initial input state of the quantum system \cite{origb7} \cite{origb8}. The “output” (result of the measurement of the witness at the final time) will change depending on the time evolution of the system, which is, of course, controlled by the Hamiltonian: specifically, by the tunneling amplitudes, the qubit biases, and the qubit-qubit coupling. Thus, we can consider these parameters to be the “weights” to be trained. We then use a quantum version \cite{origb7} of backpropagation \cite{origb13} to find optimal functions such that the desired mapping is achieved. (It should be noted that the method of quantum backprop has more recently \cite{origb14} \cite{origb15} been rediscovered by several groups.) Full details are provided in \cite{origb7}. From a training set of only four pure states, our quantum neural network successfully generalized the witness to large classes of states, mixed as well as pure \cite{origb8}. Qualitatively, what we are doing is using machine learning techniques to find a “best” hyperplane to divide separable states from entangled ones, in the Hilbert space. 

This method is adapted to form the quantum Discriminator in Fig. \ref{Quantum-GAN}.  The quantum Discriminator takes as input a density matrix representing a ``real'' or a ``fake'' state.  The output is a combination of measures made on the final time state of the quantum system.  The quantum parameters of the Discriminator Hamiltonian are what is learned to train the Discriminator to identify states as either real or fake.  A separate quantum Generator takes a flat (equal superposition of all basis states)  quantum state as its input, and is trained to generate a set of real states as its output final time state. Following the work of Karras’s Stylegan architecture for classical image GANs, the quantum Discriminator has a set of “Style” quantum parameters the modify the “common” quantum parameters in the Hamiltonian of the quantum Generator.  For each real or fake state to be generated, a classical neural network takes as input a random vector of length 6 and maps that to output a set of “Style” parameters that add to those in the Hamiltonian.  The quantum Generator internal parameters are trained on the entire set  of real data, to generate a set of “common” parameters that represent the entire real dataset.  Then the classical network is trained to create a one-to-one mapping of each member of the random pool to exactly produce a matching real in the data.  A separate random pool is then generated to produce fakes.  During the QGAN MiniMax training phase (similar to the classical GANs of Karras), both the quantum Generator, via the “Style” parameters, and the Discriminator quantum parameters are trained.  The “Style” parameters are trained to defeat the Discriminator’s ability to detect fakes, and the Discriminator quantum parameters are trained to detect these improved fakes from the Generator correctly.

Of course, this method is necessarily ``off-line'' training, since it is not possible to backpropagate without knowing the state of the system at intermediate times (in the hidden layers); quantum mechanically, this is impossible without collapsing the wavefunction and thereby destroying the superposition, which rather obviates the whole purpose of doing quantum computation. That is, quantum backpropagation can only be done on an (auxiliary) classical computer, simulating the quantum computer, and this simulation will necessarily contain uncertainties and errors in modeling the behavior of the actual quantum computer. The results from offline quantum backpropagation can, of course, be used as a good starting point for true online quantum learning, where reinforcement learning is used to correct for uncertainty, noise, and decoherence in the actual hardware of the quantum computer. In Section III we give details of Levenberg-Marquardt Machine Learning for Deep Time Quantum Networks that is used to train both the quantum Generator Hamiltonian “common” parameters and the quantum Discriminator Hamiltonian, as well as our original quantum backprop method.

\section{Levenberg-Marquardt Machine Learning for Deep Time Quantum Networks}

\subsection{Machine Learning in Simulation}\label{AA}

A general quantum state is mathematically represented by its density matrix, $\rho$, whose time evolution obeys the Schrödinger equation with a Hamiltonian $H$. The formal solution  is 
\begin{equation} \label{Liou_timeprop}
\rho(t) = e^{i L_{iou} t} \rho(t_0)
\end{equation}
% renamed Liouville operator "Liou" so it is a different symbol rom the one for the Lagrangian, and fixed the DEFINITION of Liou which was wrong. (it was  Liou = (1/hbar)[H, rho] which is wrong.}
where $L_{iou}$ is the Liouville operator, defined as the commutator with the Hamiltonian in units of  $\hbar$,  
$L_{iou}=\frac{1}{\hbar} [H,  ]$. We can think of the solution as analogous to the equation for information propagation in a neural network, Output = $F_W$*Input, where $F_W$ represents the functional action of the network as it acts on the input vector Input. The time evolution of the quantum system maps the initial state (input) to the final state (output) in much the same way. The mapping is accomplished by eqn. \ref{Liou_timeprop}. The parameters playing the role of the adjustable weights in the neural network are the parameters in the Hamiltonian that control the time evolution of the system: the physical interactions and fields in the quantum hardware, which can be specified as functions of time. Because we want to be able to implement our method physically, we use not the final state of the system, $\rho(t)$, as our output, but instead a measure, $M$,  applied to the quantum system at that final time, producing the output $O(t_f) = M(\rho(t_f))$. “Programming” this quantum computer, the act of finding the program steps (or ``algorithm'') involves determining the parameter functions that yield the desired computation. We use machine learning to find that needed quantum algorithm. This means we learn the system parameters inside $H$ so as to evolve the system in time from the initial (input) to target final (output) states; yielding a quantum system whose time evolution accurately approximates a chosen function, such as logic gates, benchmark classification problems,  a quantum function like entanglement, or image classification. If we think of the time evolution operator in terms of the Feynman path integral picture \cite{origb18}, the system can be seen as analogous to a deep neural network, yet quantum mechanical. That is, instantaneous values taken by the quantum system at intermediate times, which are integrated over, play the role of “virtual neurons” \cite{origb7}. In fact, this system is a deep learning system (Deep Time Quantum Network), as the time dimension controls the propagation of information from the input to the output of the quantum system, and the depth is controlled by how finely the parameters are allowed to vary with time. We use the term “dynamic learning” to describe the process of adjusting the parameters in this differential equation describing the quantum dynamics of the quantum computer hardware. The real time evolution of a multi-qubit system can be treated as a neural network, because its evolution is a nonlinear function of the various adjustable parameters (weights) of the Hamiltonian. 

\subsection{Quantum Levenberg-Marquardt Algorithm}

The quantum system used in this research has the parameters $K$, for the tunneling, $\epsilon$, bias, and $\zeta$, coupling coefficients, and the Hamiltonian is constructed from these parameters as:
 
\begin{equation} \label{Hamiltonian}
\displaystyle H = \sum_{i=1}^n K_{i}\sigma_{x,i} + \sum_{i=1}^n \epsilon_{i}\sigma_{z,i}+\frac{1}{2}\sum_{i\ne j}^n \zeta_{i,j}\sigma_{z,i}\sigma_{z,j}
\end{equation}

where $K_{i}$ is the tunneling coefficient for qubit $i$, $\epsilon_{i}$ is the bias coefficient for qubit $i$, and $\zeta_{i,j}$ is the coupling coefficient between the $ij$ pair of qubits in the system, and $\sigma_{x}$ and $\sigma_{z}$ are the Pauli spin matrices.  We also include optional Lindblad Markovian dynamics  as detailed below.

The parameters $\{K, \epsilon,\zeta\}$ of the Hamiltonian are time dependent, as then so is $H$.  The quantum system state variable is the density matrix, $\rho$, which evolves from an initial state $\rho_0$ at $t=0$ to $\rho^f$ at the final time $t_f$ according to the Schrodinger equation
\begin{equation} \label{steck0}
\frac{d\rho}{dt} =\frac{1}{ih}[H,\rho] + H_{lb}
\end{equation}
where $H_{lb}$ is optional Lindblad Markovian dynamics given by
\begin{equation} \label{Lindblad}
    H_{lb}=\frac{1}{\hbar}\Gamma\left(\sigma^-\rho\sigma^+-\frac{1}{2}\left\{\sigma^+\sigma^-,\rho\right\}\right)
\end{equation}
and $\Gamma$ is a time dependent quantum parameter controlling the strength of these dynamics\cite{Lindblad_Intro}. The quantum parameters are trained using machine learning to achieve a target final state $\rho^f=T$, where the initial $\rho_0$ is paired with a target final state $T$.  More details about this are given later. The time dependence of each parameter is represented as a finite Fourier series, and the Fourier coefficients are trained using an adjoint Levenberg-Marquardt (LM) machine learning method described below.

\subsection{Cost Function with Output Measures}
As stated above, the quantum system evolves in time from an initial density matrix $\rho_0$ to a final density matrix $\rho^f$ at the final time $t_f$. The quantum system obeys the Schrodinger equation which gives the dynamics of the density matrix $\rho$. The Hamiltonian, $H$, is time varying. A cost function, $L$, is defined as a positive definite error based on the Frobenius norm of a matrix.  Here, $T$ is the target output density matrix for the quantum system and $\rho$ at $t^f$ is the actual quantum system output density matrix.  The system dynamics are enforced as a constraint with Lagrange multiplier matrix $\gamma$, where $\odot$ is the element-by-element dot product of two matrices. 
\begin{multline} \label{steck1}
L = \sum (T_{ij} - \rho_{ij})^\dagger (T_{ij} - \rho_{ij}) |_{t_f} \\
+ \int_0^{t_f} \gamma^\dagger \odot \left[ \frac{d\rho}{dt} - \frac{1}{ih}[H,\rho]-H_{lb} \right] \lambda dt \\
+ \int  \left[ \frac{d\rho}{dt} - \frac{1}{ih}[H,\rho] -H_{lb}\right]^\dagger \odot \gamma dt
\end{multline}
If we use a correlation measure(s) M, as is done for the results later in this paper then the first term is replaced by
\begin{equation} \label{steck1a}
L_{1st} = \frac{1}{2}\left(\text{trace}(M \rho)^2-T_{arget}\right)^2 |_{t_f}
\end{equation}
Taking the variation with respect to $\gamma$ and setting it equal to zero enforces the Schrodinger equation dynamics. Integrating the first term in each integral by parts and taking the variation with respect to $\rho$ and setting that equal to zero at any time $t<t_f$ gives a co-state equation that is solved backward in time (``quantum error backpropagation'' )\cite{origb7}

\begin{equation} \label{steck2}
\frac{d\gamma}{dt} = -\frac{1}{ih}[H,\gamma] + L_{lb}
\end{equation}
where the optional Lindblad contribution is
\begin{multline} \label{LBbackprop}
    L_{lb}=\frac{-\Gamma}{\hbar}\sum_{kl}\lambda_k\sigma^-_{ki}\sigma^+_{jk}\gamma_l\\ 
    +\frac{\Gamma}{2\hbar}\sum_k\lambda_k(\sigma^+\sigma^-)_{ki}\gamma_j\\
    +\frac{\Gamma}{2\hbar}\sum_k\lambda_i(\sigma^+\sigma^-)_{jk}\gamma_k
\end{multline}
and its complex conjugate equation.  Note that since $\rho$ is specified at $t=0$, its variation at that initial time is zero, and that initial condition term vanishes. For a target output density matrix, at $t_f$ we have the final time condition (and its conjugate):
\begin{equation} \label{steck3}
\gamma_{ij} = (T_{ij} - \rho_{ij})
\end{equation}
If using a measure $M$, then, at the final time
\begin{equation} \label{steck2a}
\gamma_{ij} = 2\left(tr(M \rho)^2-T_{arget}\right) tr(M \rho)M
\end{equation}

The Hamiltonian, $H$, contains parameters mentioned above which describe how $H$ varies with time. Let $w$ be one of these parameters. Taking the variation of $L$ with respect to $w$ gives a gradient of $L$ that can be used in various gradient based adaptation methods to determine the parameter.

\begin{equation} \label{steck4} 
\frac{\delta L}{\delta w} = \int_0^{t_f} 2Re\left[\frac{1}{ih}\gamma^t \odot \left[\frac{dH}{dw},\rho\right]  \right]
\end{equation}
For the Lindblad parameter $\Gamma$,
\begin{equation} \label{steck4b} 
\frac{\delta L}{\delta \Gamma} = \frac{1}{\hbar}\gamma^\dagger\odot\left(\sigma^-\rho\sigma^+-\frac{1}{2}\left\{\sigma^+\sigma^-,\rho\right\}\right)
\end{equation}
% fixed this equation too- ECB
Following mainly \cite{Sam_Roweis} which is based on the Levenberg-Marquardt method of nonlinear optimization presented in \cite{K_Levenberg} \cite{D_Marquardt} and \cite{JJ_More} we derive a quasi-second order method for learning the quantum parameters in $H$.  Levenberg-Marquardt outperforms gradient descent and conjugate gradient methods for medium sized problems.

Let $\textbf{W}$ be a vector of the quantum parameters.  $L$ is a function of $\textbf{W}$, so we define $\left[\frac{\partial L}{\partial \textbf{W}}\right]$
as a vector of partials of $L$ with respect to each element of \textbf{W}
and then define $\hat{L}$ as a quadratic (2nd order) local approximation to $L$ about $\mathbf{W_0}$ the current value of the quantum parameter vector as

\begin{multline} \label{steck5} 
\hat{L}(\textbf{W}) = L(\mathbf{W_0}) + \left[\frac{\partial L}{\partial \textbf{w}}\right]^T_\mathbf{W_0} (\textbf{W}-\mathbf{W_0})\\ +\frac{1}{4L(\mathbf{W_0})}(\textbf{W}-\mathbf{W_0})\left[\frac{\partial L}{\partial \textbf{W}}\right]^T_\mathbf{W_0} \left[\frac{\partial L}{\partial \textbf{W}}\right]_\mathbf{W_0}(\textbf{W}-\mathbf{W_0})^T 
\end{multline}

Notice that $\hat{L}(\mathbf{W_0})=L(\mathbf{W_0})$
and $\left[\frac{\partial \hat{L}}{\partial \textbf{W}}\right]_\mathbf{W_0} = \left[\frac{\partial L}{\partial \textbf{W}}\right]_\mathbf{W_0}$ and

\begin{equation} \label{steck6}
L(\mathbf{W_0}) = \sum (T_{ij} - \rho_{ij})^t (T_{ij} - \rho_{ij}) |_{t_f}
\end{equation}
as the forward dynamics equation constraint being enforced means the other terms in $L$ are zero.  This means  $L(\mathbf{W_0})$ is the squared ``output'' error at the final time with the current parameters $\mathbf{W_0}$.
Then setting the variation of $L$ with respect to $\textbf{W}$ equal to 0 gives

\begin{multline} \label{steck7}
0=\left[\frac{\partial \hat{L}}{\partial\textbf{W}}\right]_\textbf{W} = \left[\frac{\partial L}{\partial \textbf{W}}\right]_\mathbf{W_0}\\ + \frac{1}{L(\mathbf{W_0})}(\textbf{W}-\mathbf{W_0})\left[\frac{\partial L}{\partial \textbf{W}}\right]^T_\mathbf{W_0}  \left[\frac{\partial L}{\partial \textbf{W}}\right]_\mathbf{W_0}
\end{multline}
Then, define a Hessian by
\begin{equation} \label{steck8} 
\textbf{H}_{ess}=\frac{1}{\sqrt{L(\mathbf{W_0})}} \left[\frac{\partial L}{\partial \textbf{W}}\right]_\mathbf{W_0}\left[\frac{\partial L}{\partial \textbf{W}}\right]^T_\mathbf{W_0} \frac{1}{\sqrt{L(\mathbf{W_0})}}
\end{equation}
giving
\begin{equation} \label{steck9}
\textbf{W} = \mathbf{W_0}-\textbf{H}_{ess}^{-1} \left[\frac{\partial L}{\partial \textbf{W}}\right]^T_\mathbf{W_0}
\end{equation}
The Levenberg-Marquardt algorithm modifies this by combining this Hessian update rule with a gradient descent update rule that is weighted by a parameter $\Lambda$
which is updated dynamically during the learning process.  Also, the Hessian and the gradient of $L$ become averages over all training pairs in the training set, or in a mini-batch subset, and $L(\mathbf{W_0})$ becomes the average squared error over the entire set, by necessity, since it could be zero for any single training data point. For details on the LM process, including updating $\Lambda$ see \cite{Steck_Behrman_Thompson}  .
The combined weight update rule is

\begin{equation} \label{steck10}
\textbf{W} = \mathbf{W_0}-\eta[\textbf{H}_{ess}+\Lambda\textbf{I}]^{-1} \left[\frac{\partial L}{\partial w}\right]^T_\mathbf{W_0}
\end{equation}

where $\textbf{I}$ is the identity matrix and $\eta$ is a constant that controls the rate of update ``learning'' of the quantum parameters.
The standard LM algorithm is
\begin{equation} \label{steck11}
\textbf{W} = \mathbf{W_0}-\eta[\hat{\textbf{H}}_{ess}+\Lambda\textbf{I}]^{-1} E \left[\frac{\partial O}{\partial w}\right]^T_\mathbf{W_0}
\end{equation}
If we absorb $L(\textbf{W}_0)$ into $\eta$ and $\Lambda$ and define the output $O$ in the standard LM algorithm above as the squared error $L$, and have its target value as zero, our update using $L$ looks much like the standard LM formulation. Because the error of our method at the final time 
\begin{equation} \label{steck12} 
\gamma_{ij} = (T_{ij} - \rho_{ij}) 
\end{equation} 
is propagated backward through the quantum system dynamics in equation \ref{steck2}, the error is contained in the gradient calculated via equation \ref{steck4}.  Therefore the ``error'' $E$ which multiplies the gradient in the standard LM algorithm is set to a vector of ones in our LM formulation.

The motivation for the approximation $\hat{L}$ is as follows. Let a function of one variable $x$, $f(x)$, be approximated close to 0 by the form 
\begin{equation} \label{steck13}
\hat{f(x)}=\frac{1}{2}a(1+bx)^2
\end{equation}
Note this form matches the form of the first term of $L$ above.  Then choose $a=2f(0)$ to make $f(0)=\hat{f}(0)$ and choose $ab=\left[\frac{\partial f}{\partial x}\right]_0$ to make $\left[\frac{\partial f}{\partial x}\right]_0=\left[\frac{\partial \hat{f}}{\partial x}\right]_0$ .  Then

\begin{equation} \label{steck14}
\hat{f}(x) = f(0) + \left[\frac{\partial f}{\partial x}\right]_0 x +\frac{1}{4f(0)}x^2\left[\frac{\partial L}{\partial x}\right]^T_0 \left[\frac{\partial f}{\partial x}\right]_0 
\end{equation}

For small $\Lambda$, the update rule is similar to the Gauss-Newton algorithm, allowing larger steps when the error is decreasing rapidly. For larger $\Lambda$, the algorithm pivots to be closer to gradient descent and makes smaller updates to the weights. This flexibility is the key to LM’s efficacy, changing $\Lambda$ to adapt the step size to respond to the needs of convergence: moving quickly through the parameter space where the error function is steep and slowly when near an error plateau and thereby finding small improvements. Our implementation follows the description from our previous paper \cite{Steck_Behrman_Thompson} and is a modified LM algorithm following several suggestions in \cite{origb24}.  Quoting from \cite{Steck_Behrman_Thompson} "One epoch of training consists of the following:
\begin{enumerate}
  \item Compute the Jacobian (13) with current weights $w$.
  \item Update the damping parameter $\Lambda$.
  \item Calculate a potential parameter update.
  \item Find if RMS error has decreased with new parameters, or if an acceptable uphill step is found.
  \item If neither condition in step 3 is satisfied, reject the update, increase $\Lambda$, and return to step 2.
  \item For an accepted parameter change, keep the new parameters and decrease $\Lambda$, ending the epoch. 
\end{enumerate}
The identity matrix $I$ that multiplies $\Lambda$ can be replaced by a scaling matrix $D^T\!D$ which serves the primary purpose of combating parameter evaporation \cite{origb26}, which is the tendency of the algorithm to push values to infinity while somewhat lost in the parameter space. Following \cite{origb24}, we can choose $D^T\!D$ to be a diagonal matrix with entries equal to the largest diagonal entries of $\textbf{H}_{ess}$ yet encountered in the algorithm, with a minimum value of $10^{-6}$. Updates to the damping factor may be done directly or indirectly; our results here use a direct method."
\section{Quantum Product States GAN - Matlab Simulation}
We begin by demonstrating that our GAN is truly quantum mechanical by applying it to an inherently quantum problem: that of entanglement classification. 

We start with a 2-qubit system, the smallest quantum system that can exhibit entanglement. (In future work, we will use quantum transfer learning \cite{origb37} to try to generalize our results to the unsolved case of higher numbers of qubits.) As explained above, we identify the "reals" as the set of product states. The set of fakes are impostor product states, i.e., partially or fully entangled.  The Discriminator tries to correctly identify reals and fakes and the Generator tries to generate fakes that successfully fool the Discriminator.  For the Discriminator, the output error needed for the LM algorithm  is a quantum measure on the final time state.  Three measures are used, each a correlation with 2-qubit Pauli state $\sigma_{xx},$ $\sigma_{yy}$ and $\sigma_{zz}$.  This creates 3 hyperplanes that separate and classify the final state into two categories.  If the measure is below a specified threshold, the state is classified as a product state, above and it is not.  For a real state to be correctly classified, it must be contained within the space bounded by all three hyperplanes, thus all three planes must classify it as a real product state. A fake is detected correctly if the final state is outside of ONLY ONE of the hyperplanes, that is only one of the correlation measures needs to be above the threshold. For an initial training of the Generator to reproduce the set of real states, the Frobenius norm is used, a measure of the comparison of the output state with the target real state.  During the GAN MiniMax training, the Generator is trained on fake states from the Generator, by propagating the error gradient at the output of the Discriminator, back through the Discriminator to obtain the Jacobian for the Generator and standard error gradients for training the classical 'style' network. For this training phase on the fakes from the Discriminator, the error of the Discriminator output is defined as to force the Discriminator to incorrectly identify these fakes as reals.  The Hamiltonian parameters K, $\epsilon$, $\zeta$ are trained.  The Lindblad parameter $\Gamma$ is set to zero for the product state quantum GAN. 

\begin{figure}[htbp]
\centerline{\includegraphics[width=1\linewidth]{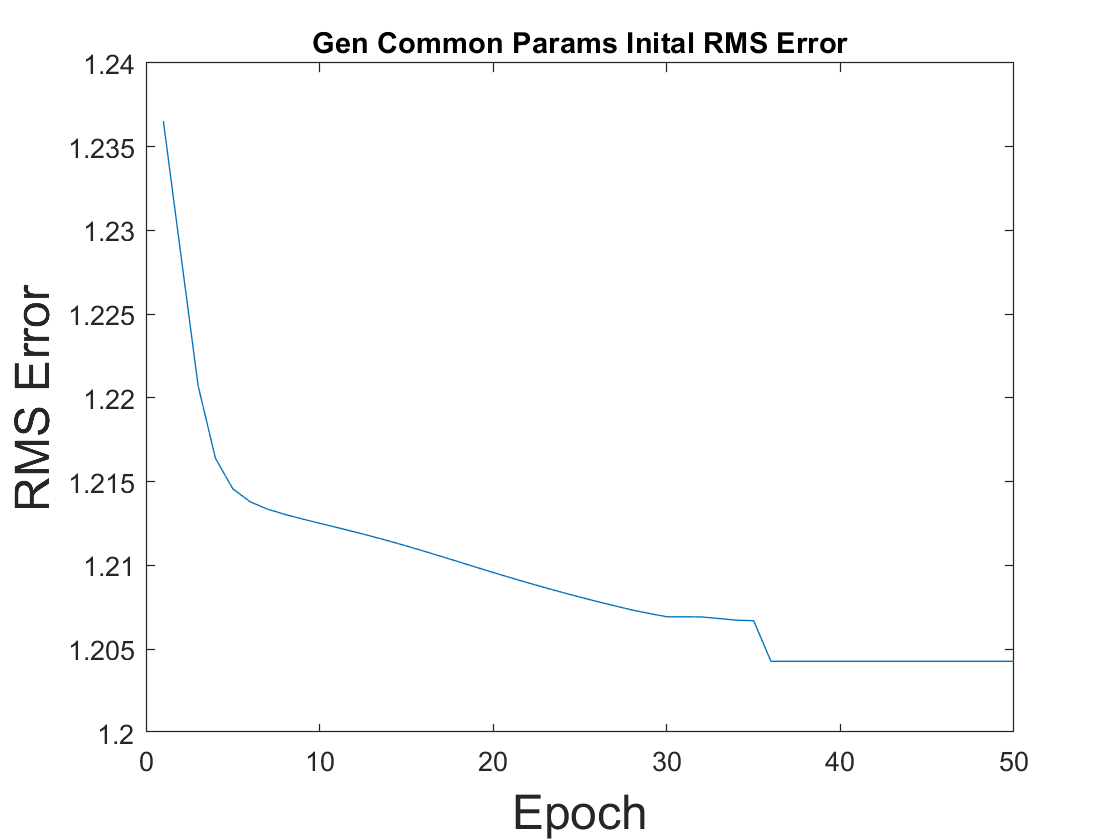}}
\caption{RMS error vs epoch for the common Generator parameters when trained on the set of real product states for 2-qubits using the Levenberg-Marquardt method in a MATLAB Simulation.}
\label{Common-Generator}
\end{figure}

The algorithm, with plots of results for 2-qubits with 80 real product states and 80 fake product states, after each step is detailed below:

1)	Initially train the Generator Hamiltonian parameters to reproduce the set of reals as a whole.  These are referred to as “common” parameters as they are common to the entire real set. The RMS error is shown in Fig. \ref{Common-Generator}.

\begin{figure}[htbp]
\centerline{\includegraphics[width=1\linewidth]{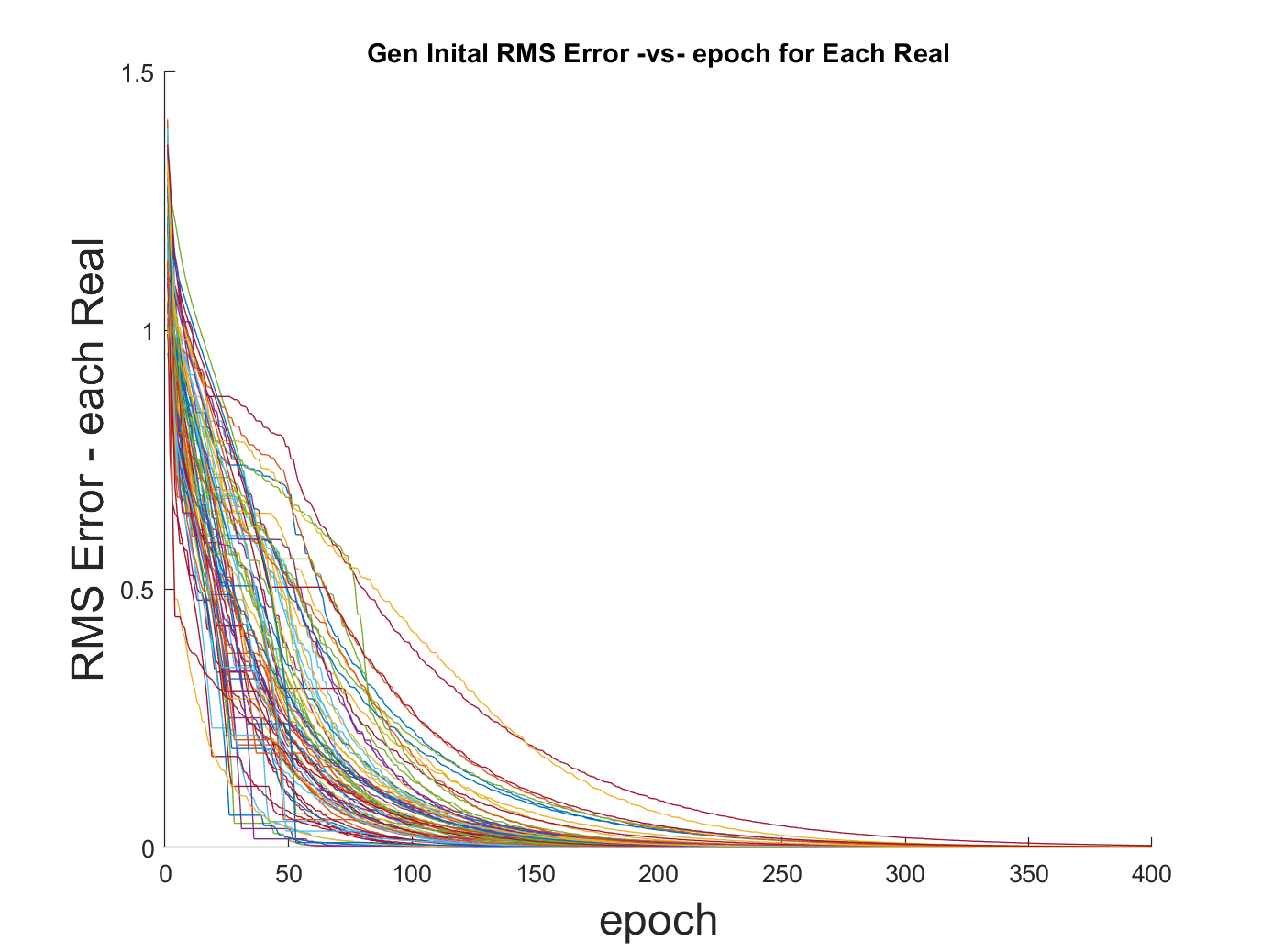}}
\caption{RMS error vs epoch during the style network training for EACH real for 2-qubits using the Levenberg-Marquardt method in a MATLAB Simulation.}
\label{Style-Generator}
\end{figure}

2)	Initially train the Generator “style” network parameters for each individual real in the set of reals.  The style network takes a vector from the random pool and one-to-one maps it to generate parameters that modify the common parameters and will accurately reproduce a specific real from that set of product states. RMS traces for each real are shown in Fig. \ref{Style-Generator}

\begin{figure}[htbp]
\centerline{\includegraphics[width=1\linewidth]{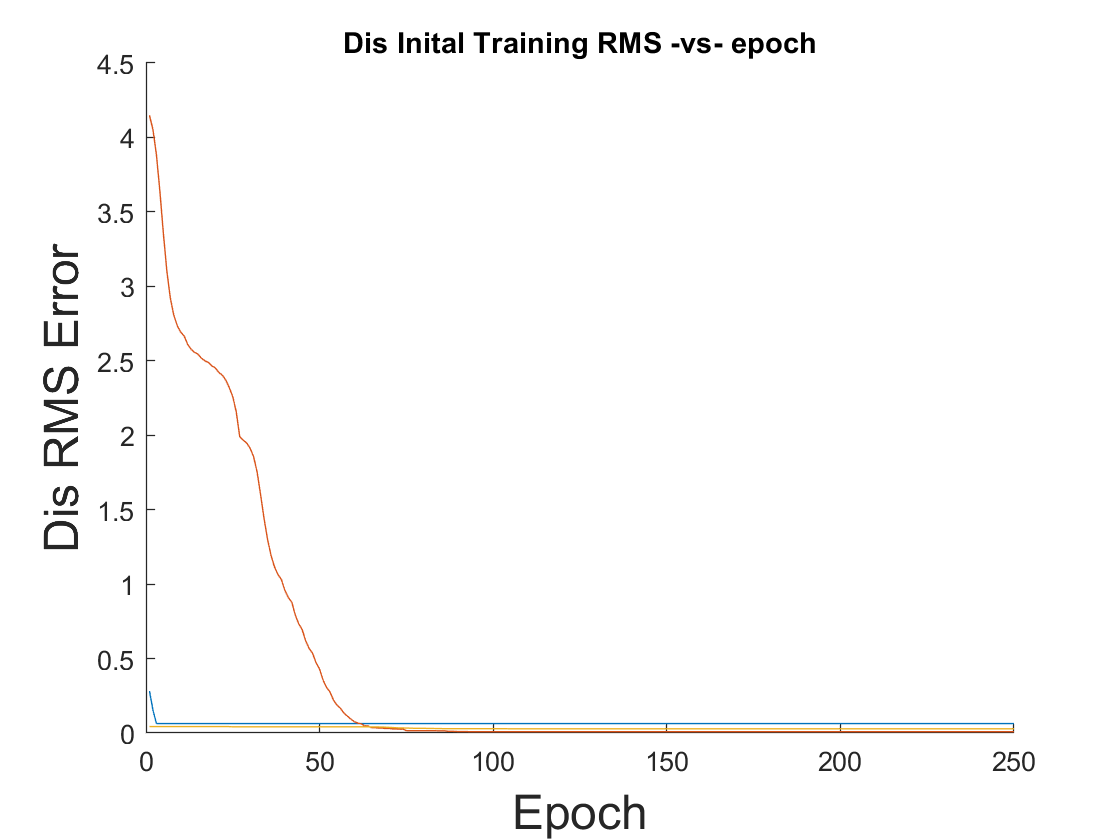}}
\caption{RMS error vs epoch for initial training of the Discriminator on the set of real product states. All 3 planes are shown (red, blue, yellow) and trained separately with errors based on that all three measures (hyperplanes) must identify the state as real for correct identification.}
\label{Initial-Discriminator}
\end{figure}

3)	Initially train the Discriminator Hamiltonian parameters to correctly identify each real product state in the set of reals used above to train the Generator. The errors are based on the fact that all three measures (hyperplanes) must identify the state as real for correct identification.  Measures that ID incorrectly are trained. RMS for the initial training of the Discriminator on the reals is shown in Fig. \ref{Initial-Discriminator}.

\begin{figure}[htbp]
\centerline{\includegraphics[width=1\linewidth]{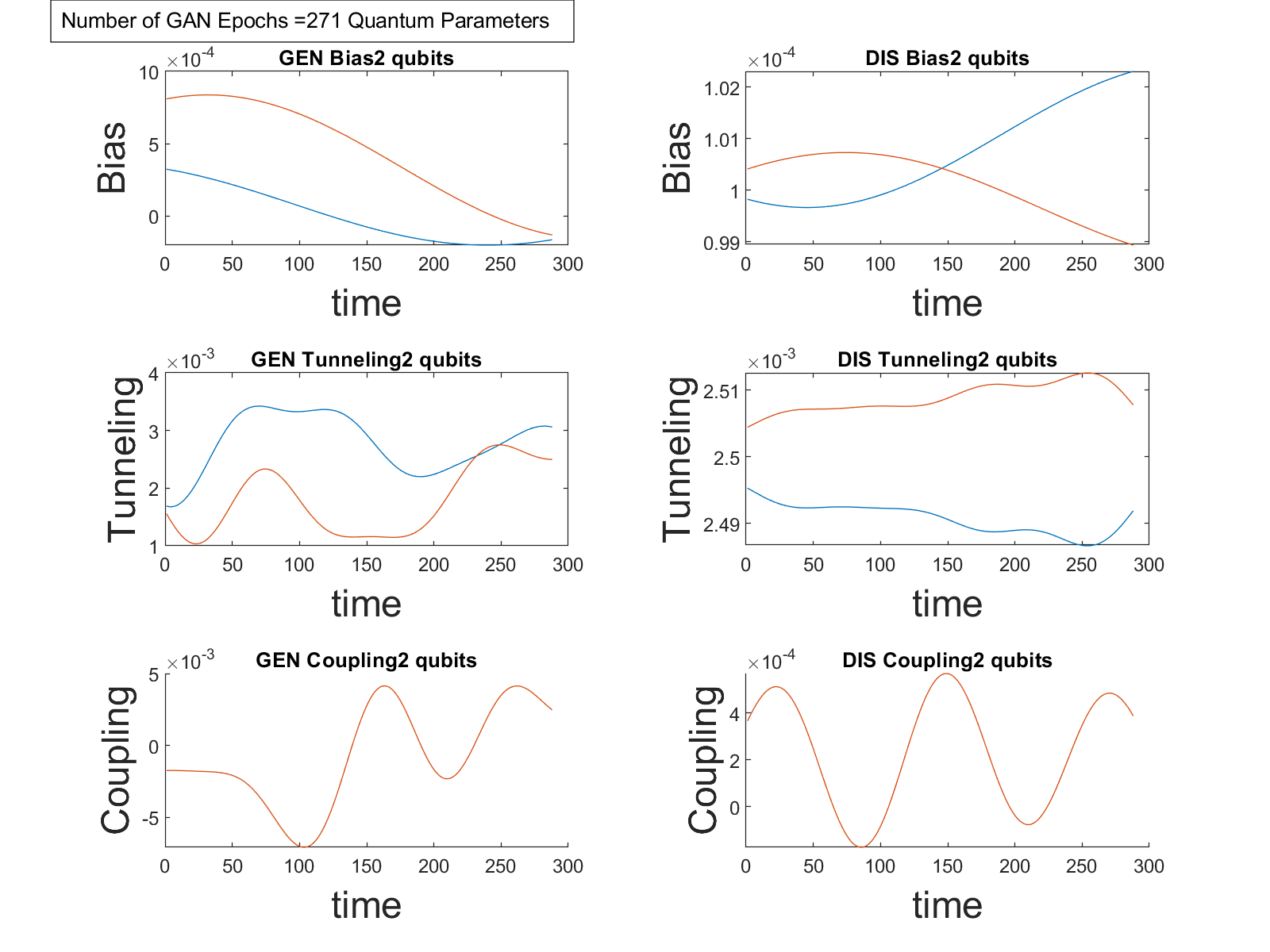}}
\caption{Quantum parameters for the Generator and Discriminator after the initial training phase.}
\label{Initial Quantum Parameters}
\end{figure}

Figure A1 at the end of the paper shows the fake product state density matrices generated by the Generator after initial training on the real states BUT BEFORE any GAN training of the Generator or Discriminator.  The green labels above each plot indicate that the Discriminator was successfully trained to correctly ID most of these during the initial training of the Generator. Red indicates incorrect Discriminator identification. Plots of the quantum parameters for the Generator (style parameters are from only the first real) and for the Discriminator are shown in Fig. \ref{Initial Quantum Parameters} after the above initial training steps.  

4)	Enter a GAN training loop that
a)	Trains the Discriminator quantum parameters to correctly identify the fake product states being generated by the Generator.  Each fake is generated based on “style” modifications from the classical network that maps  a random vector from the fake pool, one-to-one, to its style mod. Only one Discriminator measure (hyperplane) must detect a fake for it to be correctly identified.
b)	Trains the Generator ``style'' parameters for each fake generated from that same fake pool based on the reverse of the error of the Discriminator identifying it as fake (correct for the Discriminator, but an error for the Generator) or real (incorrect for the Discriminator, but not an error for the Generator).  Again, all three measures (hyperplanes) must identify a state as real for the Discriminator error to be correct.  Only one measure (hyperplane) need identify a state a fake in order for the Discriminator to be correct and therefore the Generator in this step needs to adapt to oppose that correct Discriminator identification.

\begin{figure}[htbp]
\centerline{\includegraphics[width=1\linewidth]{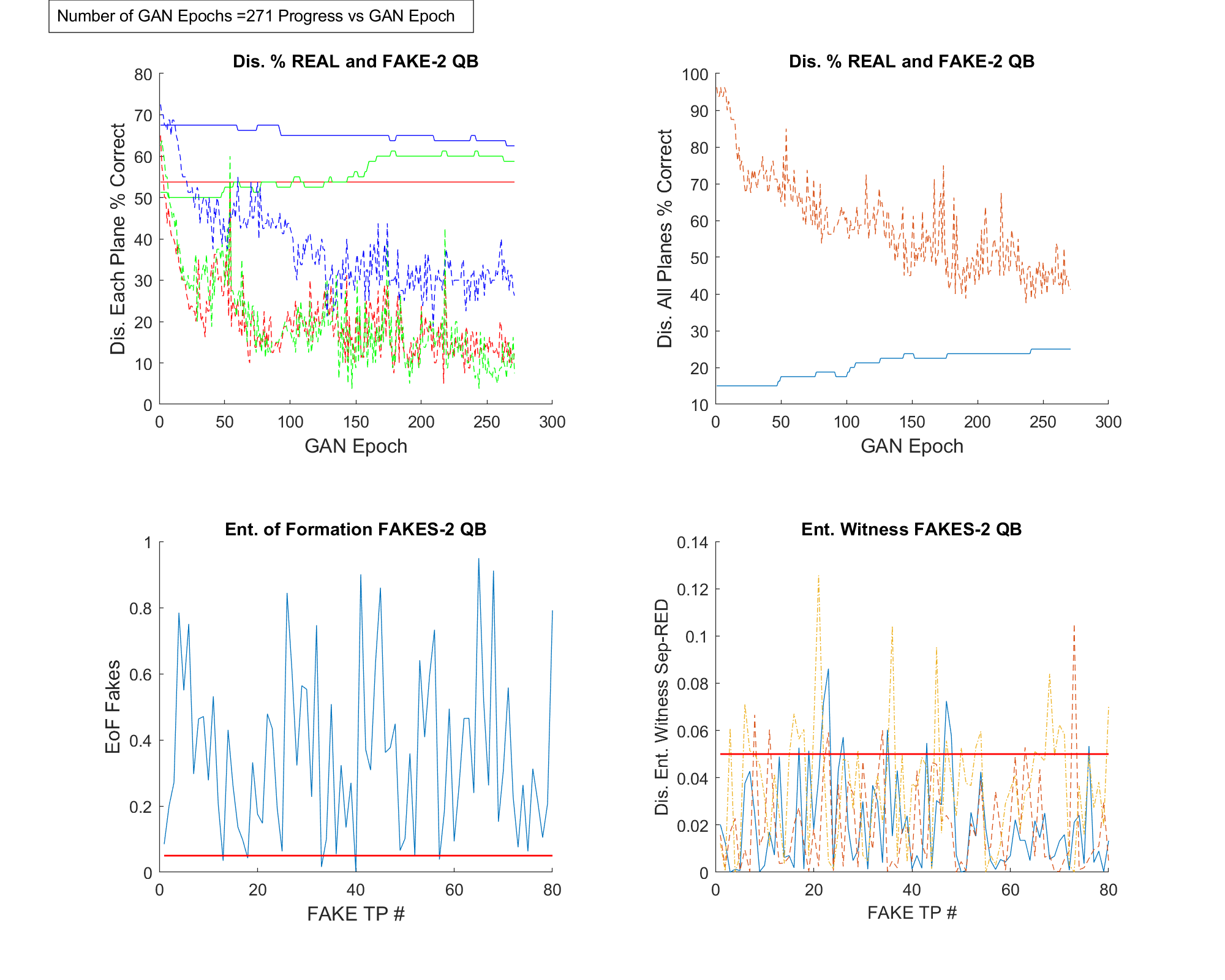}}
\caption{Quantum parameters for the Generator and Discriminator after the initial training phase.}
\label{2 qubit GAN training RMS}
\end{figure}

A plot of this GAN training progress as a function of GAN epoch is presented in Fig. \ref{2 qubit GAN training RMS} showing percent correct identification of the reals and fakes.  The top left subplot shows the percent correct identification of each of the 3 measures (hyperplanes).  The reals are solid lines and the fakes are the dashed lines.  The top left subplot shows the percent correct as an aggregate of all 3 measures (hyperplanes) where a real is identified correctly ONLY if all 3 measures of the Discriminator agree, whereas only one measure must correctly identify a fake for the Discriminator to correctly identify in aggregate.  Notice in this plot that the Generator is clearly training toward a Nash equilibrium as it is gradually forcing the Discriminator to incorrectly identify more fakes as GAN training progresses. The GAN loop only trains using the fake product states, so the real identification is static.  Figure A2 at the end of the paper shows the fake product state density matrices generated by the Generator after GAN training of the Generator and Discriminator.  The green labels above each plot indicate that the Discriminator now correctly identifies a bit more than half of these as the Generator has improved its ability to generate fake product states. Red indicates incorrect Discriminator identification.

\section{Image Binary Classification - Matlab Simulation}
The Discriminator in a GAN is a binary classifier that classifies data as real (provided data) or fake (data produced by the Generator).  As part of the development of the quantum GAN, we designed and validated the quantum Discriminator's ability to do a binary classification of the product states.  Here, using only one measure (hyperplane) we modify the quantum discriminator to be a quantum binary classifier of of  $2^n$x$2^n$ grayscale pixel images we pre-process to be represented as density matrix in a n-qubit system using a quantum image transform process as follows.
\begin{enumerate}
    \item Down-sample the image file to be a $2^n$x$2^n$ grayscale pixel image $I_{smp}$.
    \item Perform a 2D discrete Fourier Transform on the sampled image $I_{smp}$.  This results in a complex matrix $I_{FFT}$ that contains many complex conjugate pairs but that are not in the correct off locations in the matrix for it to be Hermitian.  It also may contain real off diagonal entries and complex entries on the diagonal.
    \item Rearrange the matrix $I_{FFT}$ from step 2 by swapping the real non-diagonal elements with complex elements on the diagonal.  If there are remaining complex elements on the diagonal, find off-diagonal elements with the smallest real parts (smallest spatial phase information), and swap enough of these to replace the complex elements on the diagonal.  
    \item Now make the resulting matrix to be complex conjugate symmetric by swapping complex conjugate pairs in the off diagonal positions to be in the correct symmetrical location, resulting in the matrix $I_{CCsym}$. This is done by a brute force search of each element of the matrix to find matching off diagonal complex conjugate pairs to swap.
    \item Transform the matrix $I_{CCsym}$ from step 2 to be a proper density matrix (Hermitian, trace 1 and positive semi definite)
    \begin{enumerate}
        \item $I_{PSD} = I_{CCsym}^\dagger I_{CCsym}$ 
        \vspace{2pt}
        \item $I_{DEN} = \frac{I_{PSD}}{abs(trace(I_{PSD}))}$
    \end{enumerate}
\end{enumerate}
\vspace{2pt}

\subsection{Letter Classification}
\begin{figure}[htbp]
    \centering
    \includegraphics[width=1\linewidth]{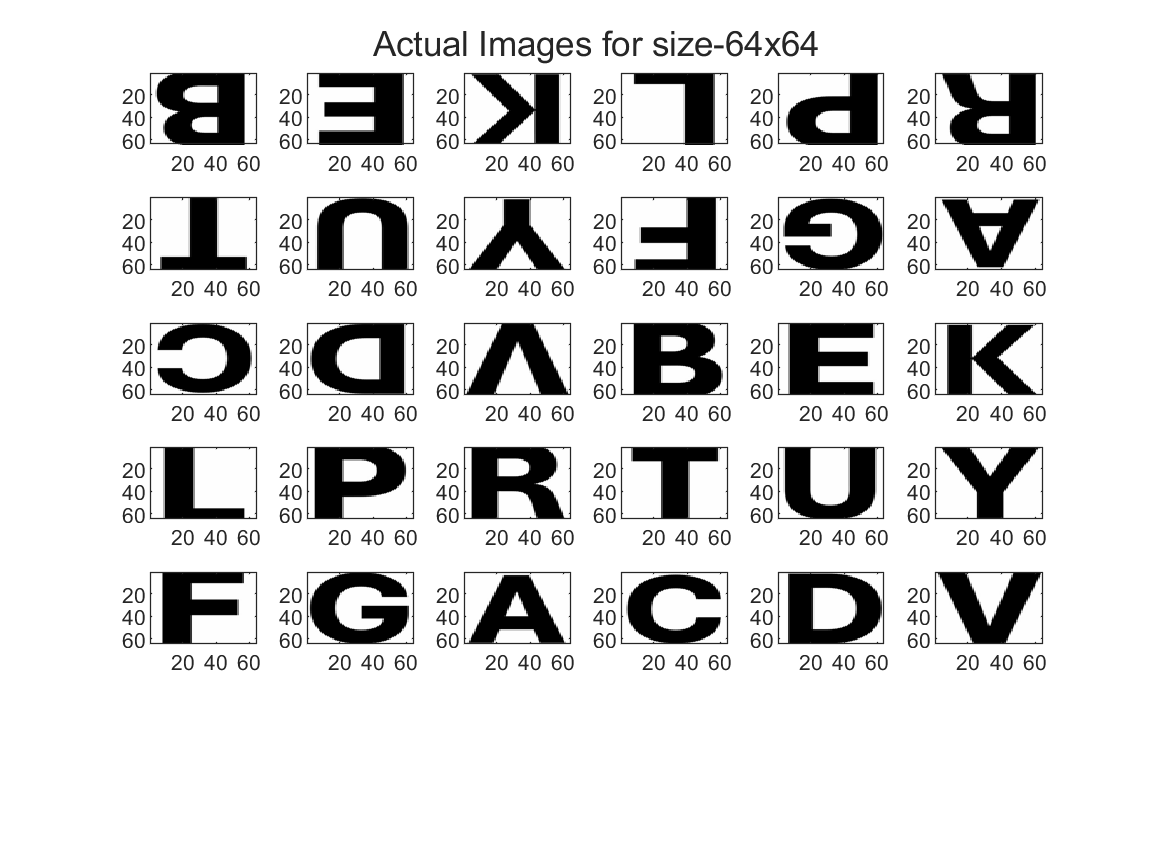}
    \caption{Letter Classification Images.}
    \label{15rightside_15upsidedown_64x64Actual_Images}
\end{figure}

A simple classification problem was created for a demonstration. 15 letter images and 15 inversions of these images were generated and saved as .jpg files in an image database.  They are shown in Fig. \ref{15rightside_15upsidedown_64x64Actual_Images}.

Each image is transformed into a n-qubit ( $2^n$x$2^n$ ) density matrix using the quantum transform procedure above. Each density matrix then becomes the ``input'' to the Discriminator/classifier by assigning the initial state of the quantum Discriminator at $t_0$ to be that density matrix.  The Discriminator quantum dynamics propagate that initial state to a final state at $t_f$.  A measure of the correlation of that final state to the n-qubit Pauli state $\sigma_{zz}$ is made.  This becomes the output of the Discriminator/classifier.  This measure value is compared to a binary separator value to achieve the binary classification of ``diagonal left'' (below the separator) and ``diagonal right'' (above the separator).  A target output is defined to be above, or below, the separator, depending upon what is the correct classification.  The Hamiltonian parameters $K$, $\epsilon$, $\zeta$ and the Lindblad parameter $\Gamma$ are trained.  
\subsubsection{3 qubit 8x8 Image Results}
\begin{figure}
    \centering
    \includegraphics[width=0.75\linewidth]{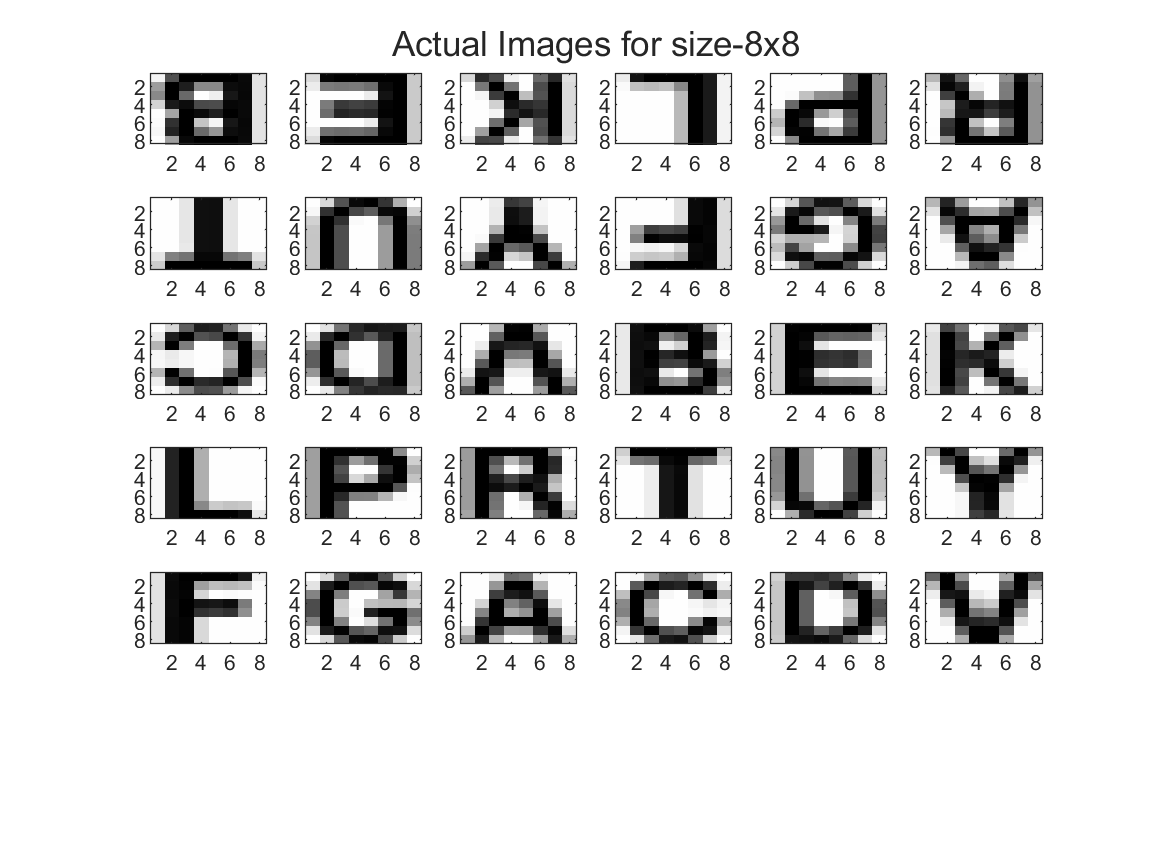}
    \caption{3 qubit 8x8 Pixel Letter Images}
    \label{3_qubit_8x8Letters}
\end{figure}

\begin{figure}[htbp]
\centerline{\includegraphics[width=1\linewidth]{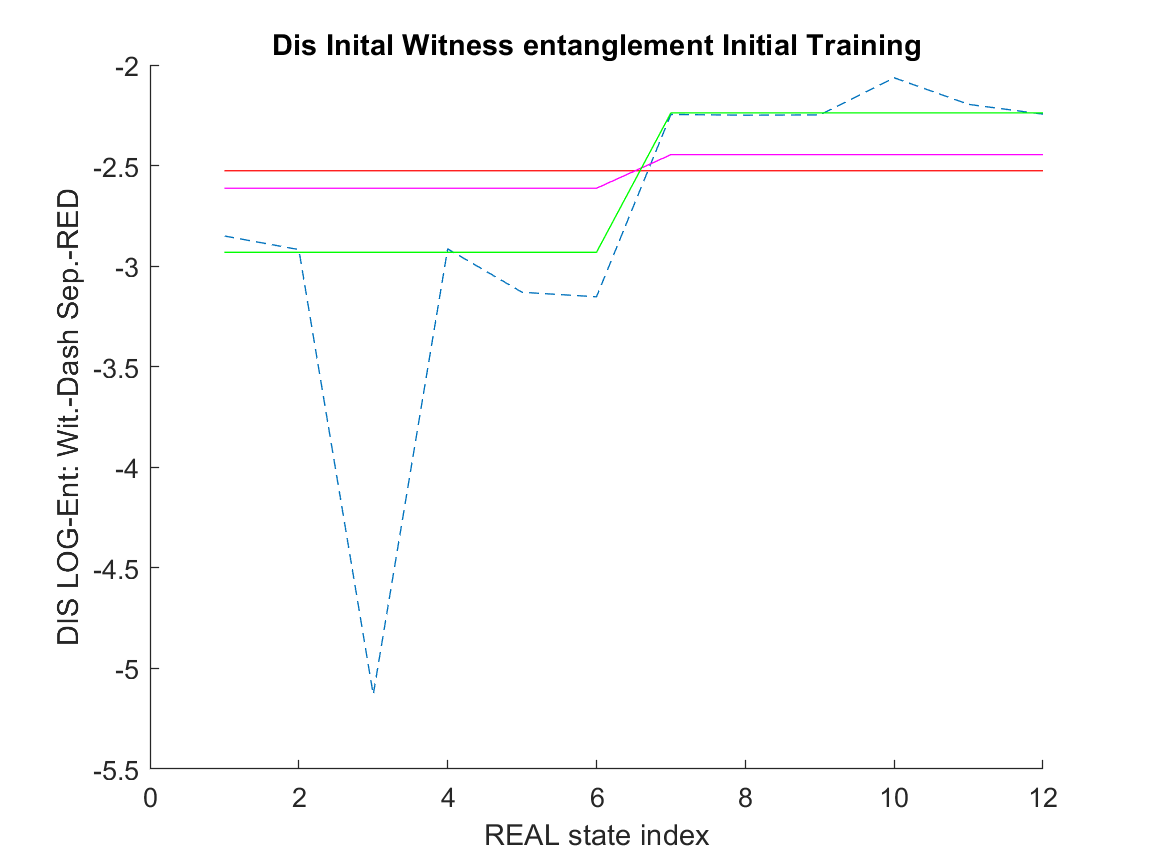}}
\caption{3 qubit Discriminator Training Classification. Red line is separator, Green line is Discriminator target measure values, Magenta line is 1/4 the distance to the target green line, Blue Dashed line is actual measure values.}
\label{3_qubit_6_Letters_Training_Results}
\end{figure}

\begin{figure}[htbp]
\centerline{\includegraphics[width=1\linewidth]{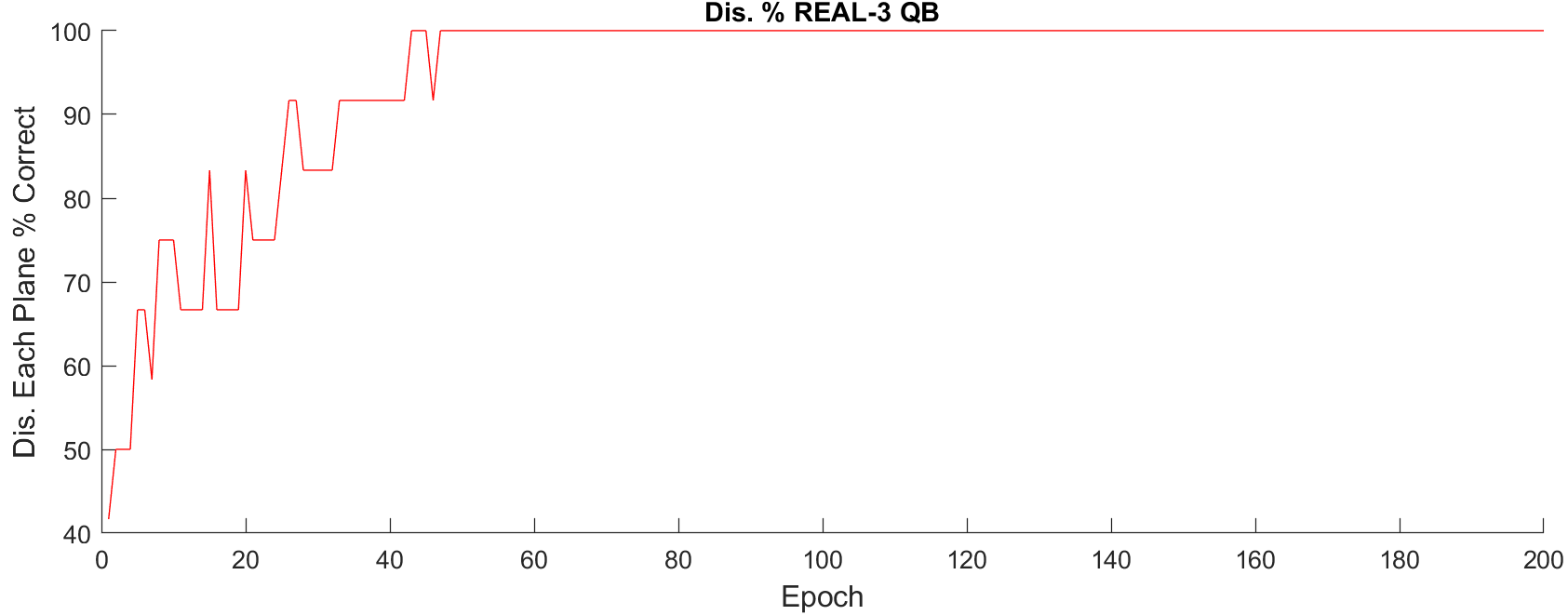}}
\caption{Discriminator Training RMS -vs- epoch.}
\label{3_qubit_6_Letters_Classification_Results}
\end{figure}

Downsampling all the letter images to a very coarse 8x8 pixel resolution gives the 15 inverted and 15 non-inverted letter images in Fig. \ref{3_qubit_8x8Letters}
Following quantum image transformation, the discriminator is tasked with classifying 6 inverted and 6 non-inverted images.  Training results are shown  in Fig. \ref{3_qubit_6_Letters_Training_Results}. The separator value is shown as the red line and the magenta line is 1/4 the distance from the separator to the target green line.The target values for the measured Discriminator output are shown as the green line  and the after training Discriminator output measure values are shown as the blue dotted line. During training, if an input is classified at or beyond the green target value the error used for training is discounted by a factor of 100 as this input is clearly correctly classified.  If an input is classified between the green target value and the magenta value, the error used for training is discounted by a factor of 5 as this input is  correctly, but not clearly classified. if an input is classified on the side of the magenta line nearest the separator, the error used for training is not discounted at all as this input is near enough to being NOT correctly classified. The $x$-axis is the image pair number.  Inverted images are numbers 1-6 and non-inverted images are numbers 7-12.  The trained 100 percent classification result is shown in Fig. \ref{3_qubit_6_Letters_Classification_Results}.

\subsubsection{4 qubit 16x16 Image Results}
\begin{figure}
    \centering
    \includegraphics[width=0.75\linewidth]{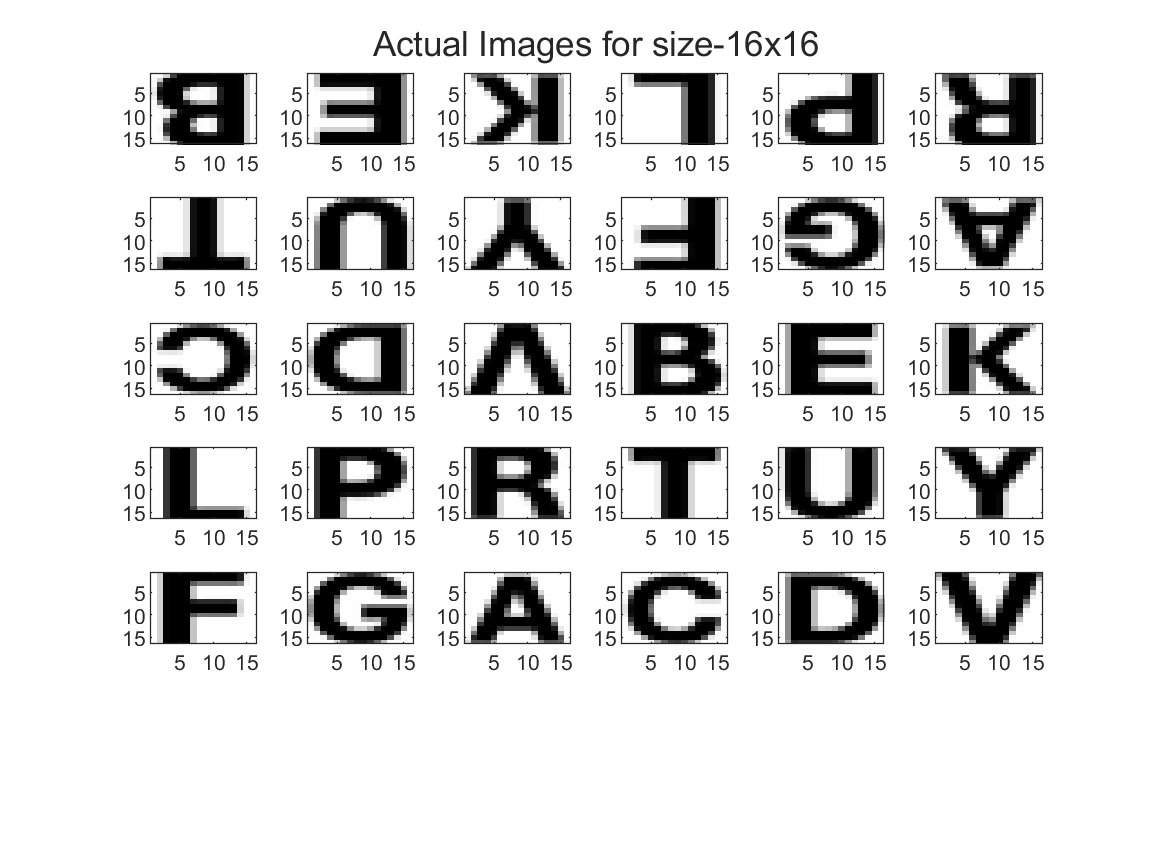}
    \caption{4 qubit 16x16 Pixel Letter Images}
    \label{3_qubit_16x16Letters}
\end{figure}

\begin{figure}[htbp]
\centerline{\includegraphics[width=1\linewidth]{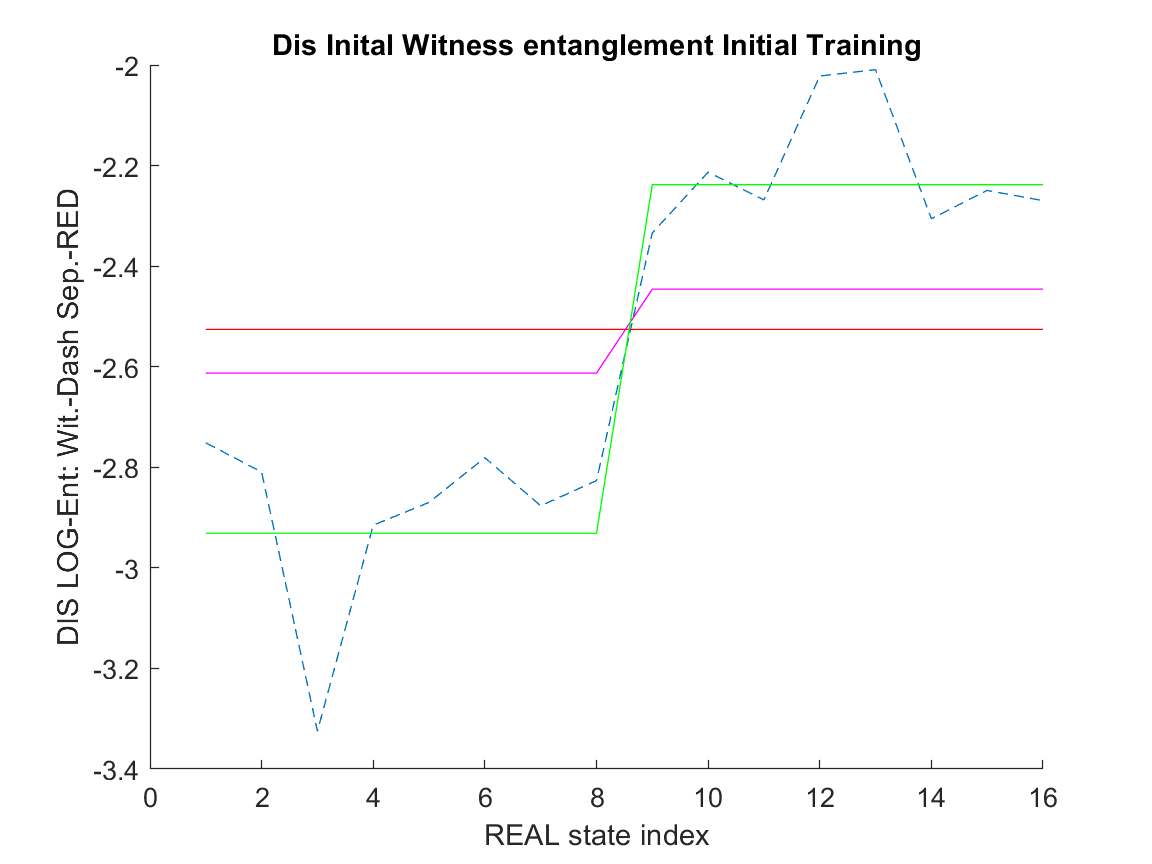}}
\caption{3 qubit Discriminator Training Classification. Red line is separator, Green line is Discriminator target measure values, Magenta line is 1/4 the distance to the target green line, Blue Dashed line is actual measure values.}
\label{8down_8up_4qubits_Training_Results}
\end{figure}

When 2 additional letters are added to the classification task, the 3 qubit system is unable to correctly classify one of the letters.  Increasing the resolution to 4 qubits, or 16x16 pixel resolution improves that result.  The 15 letters at 16x16 pixel resolution are shown in Fig. \ref{3_qubit_16x16Letters}, the Discriminator classification result is shown in Fig. \ref{8down_8up_4qubits_Training_Results}

\begin{figure}[htbp]
\centerline{\includegraphics[width=1\linewidth]{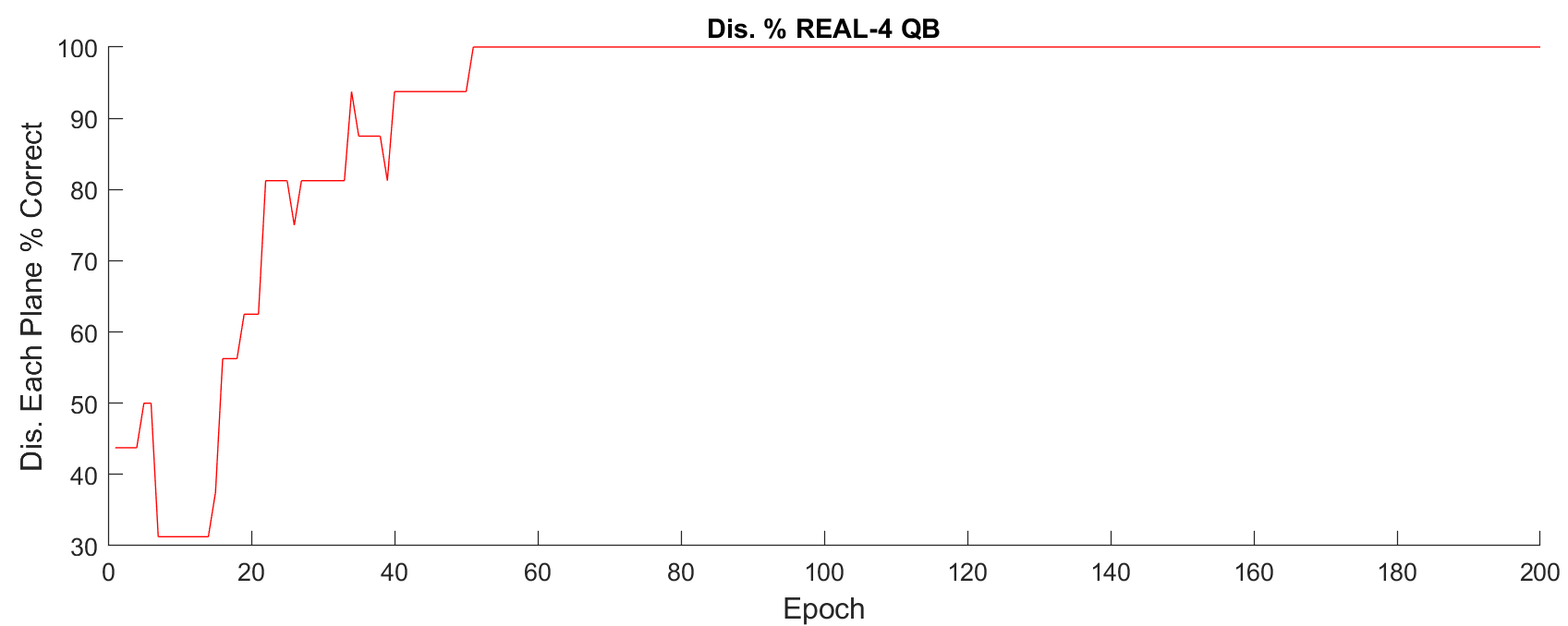}}
\caption{Discriminator Training RMS -vs- epoch.}
\label{4_qubit_8_Letters_Classification_Results}
\end{figure}

The trained 100 percent classification result is shown in Fig. \ref{4_qubit_8_Letters_Classification_Results}.

Adding more letters for a total of 12 in each class results in a 75 percent correct classification. 

\subsection{Cats -vs- Birds Discrimination}
\begin{figure}[htbp]
    \centering
    \includegraphics[width=1\linewidth]{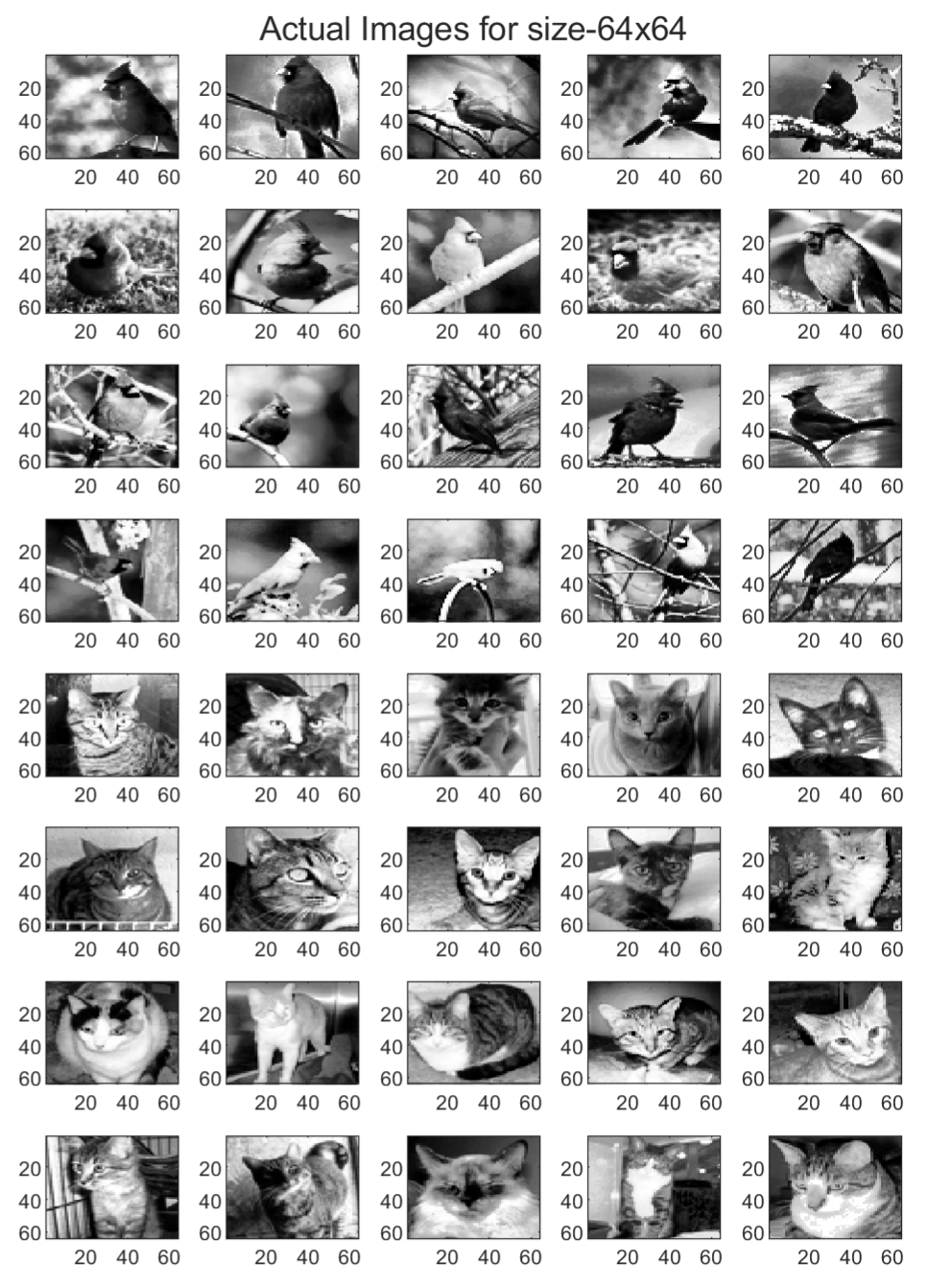}
    \caption{Bird and Cat Classification Images.}
    \label{20birds20cats_64x64Actual_Images_cropped}
\end{figure}

Another more challenging classification problem uses images of cats and birds as .jpg files in image databases \cite{bird_database} \cite{dogandcat_database}.  Twenty of each are shown in Fig. \ref{20birds20cats_64x64Actual_Images_cropped}.

As with the the letter classification application above, each image is transformed into a n-qubit ( $2^n$x$2^n$ ) density matrix using the quantum transform procedure above. Each density matrix then becomes the ``input'' to the Discriminator/classifier by assigning the initial state of the quantum Discriminator at $t_0$ to be that density matrix.  The Discriminator quantum dynamics propagate that initial state to a final state at $t_f$.  A measure of the correlation of that final state to the n-qubit Pauli state $\sigma_{zz}$ is made.  This becomes the output of the Discriminator/classifier.  This measure value is compared to a binary separator value to achieve the binary classification of ``diagonal left'' (below the separator) and ``diagonal right'' (above the separator).  A target output is defined to be above, or below, the separator, depending upon what is the correct classification.  The Hamiltonian parameters $K$, $\epsilon$, $\zeta$ and the Lindblad parameter $\Gamma$ are trained.  
\subsubsection{4 qubit 16x16 Image Results}
\begin{figure}
    \centering
    \includegraphics[width=1\linewidth]{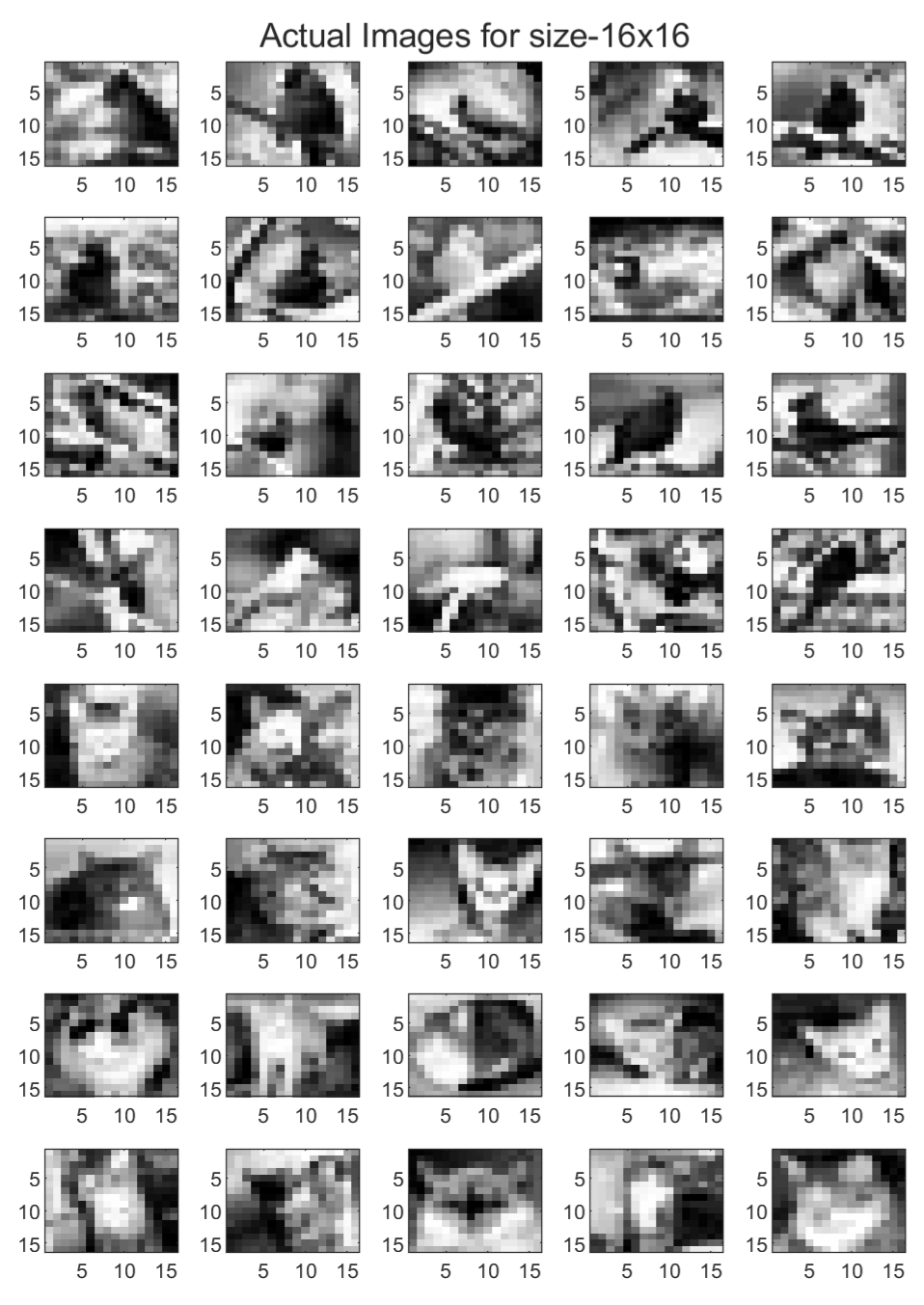}
    \caption{4 qubit 16x16 Pixel Bird and Cat Images}
    \label{20birds20cats_16x16Actual_Images_cropped}
\end{figure}

\begin{figure}[htbp]
\centerline{\includegraphics[width=1\linewidth]{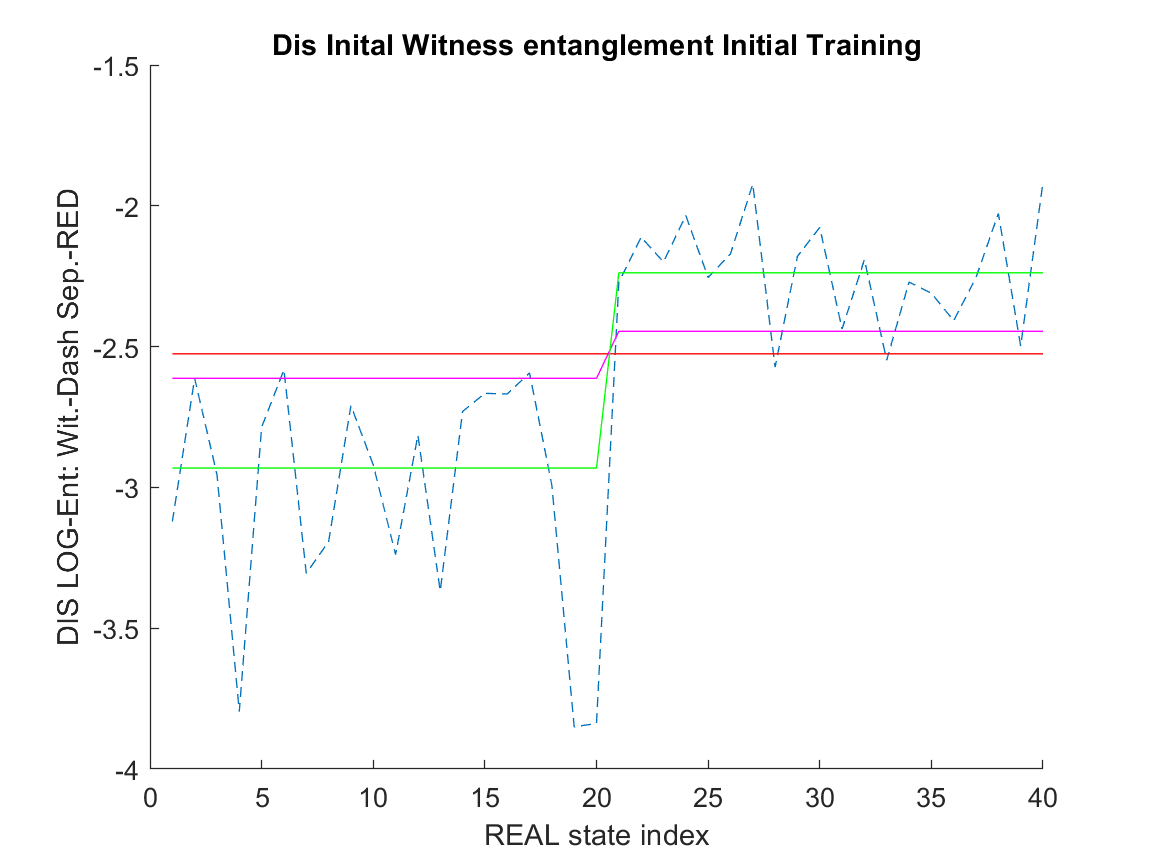}}
\caption{2 qubit Discriminator Training Classification. Red line is separator, Green line is Discriminator target measure values, Magenta line is 1/4 the distance to the target green line, Blue Dashed line is actual measure values.}
\label{20birds20cats_4qubits_Training_Results}
\end{figure}

\begin{figure}[htbp]
\centerline{\includegraphics[width=1\linewidth]{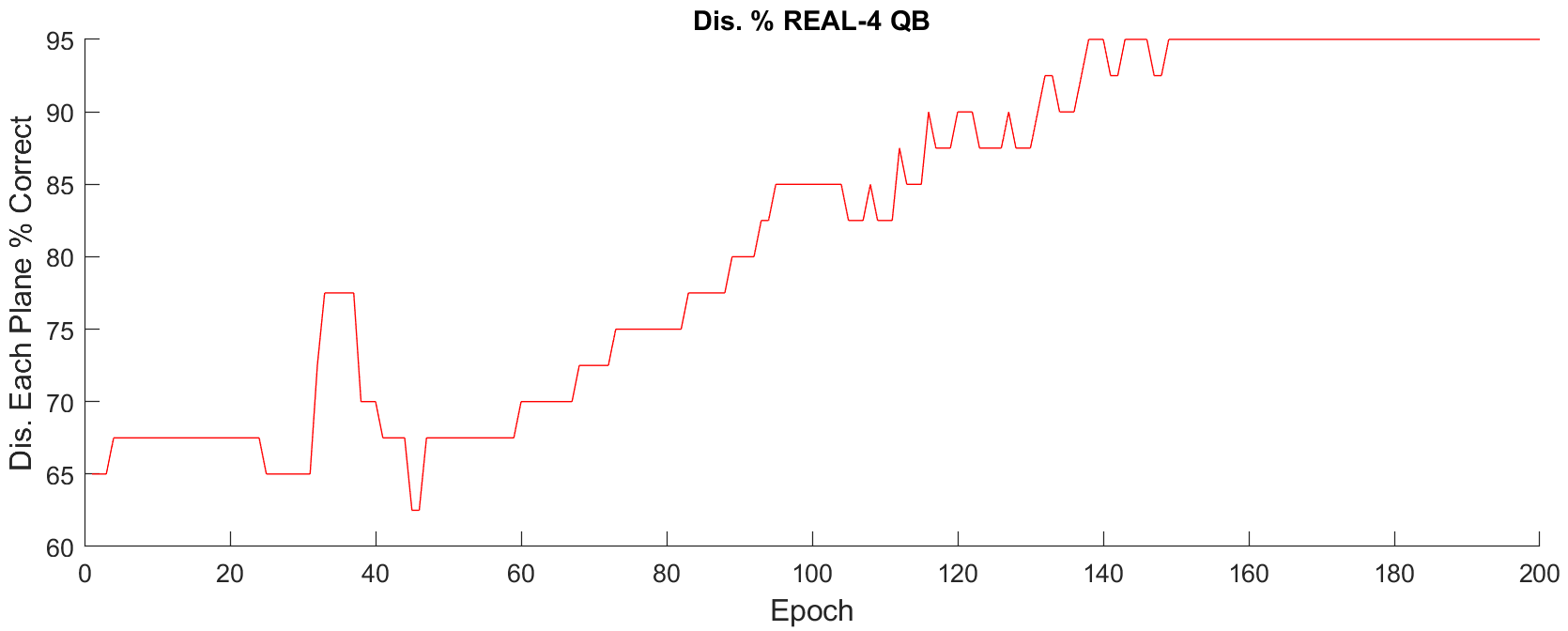}}
\caption{Discriminator Training Results Percent Classification -vs- Epoch.}
\label{20birds20cats_4qubits_Percent_Classification}
\end{figure}

Down sampling the letter images for recognition with 4 qubits gives a very coarse 16x16 pixel resolution.  The down sampled images used, twenty for each of cats and dogs, are shown in Fig. \ref{20birds20cats_16x16Actual_Images_cropped}
Following quantum image transformation, the discriminator is tasked with classifying them.  Training results are shown in Fig. \ref{20birds20cats_4qubits_Training_Results}. The separator value is shown as the red line and the magenta line is 1/4 the distance from the separator to the target green line.The target values for the measured Discriminator output are shown as the green line  and the after training Discriminator output measure values are shown as the blue dotted line. During training, if an input is classified at or beyond the green target value the error used for training is discounted by a factor of 100 as this input is clearly correctly classified.  If an input is classified between the green target value and the magenta value, the error used for training is discounted by a factor of 5 as this input is  correctly, but not clearly classified. if an input is classified on the side of the magenta line nearest the separator, the error used for training is not discounted at all as this input is near enough to being NOT correctly classified.  The $x$-axis is the image pair number.  Birds are numbers 1-20 and cats are numbers 21-40. A plot of the RMS error during Discriminator training is shown in Fig. \ref{20birds20cats_4qubits_Percent_Classification}.

\subsection{Cats -vs- Dogs Discrimination}
\begin{figure}[htbp]
    \centering
    \includegraphics[width=1\linewidth]{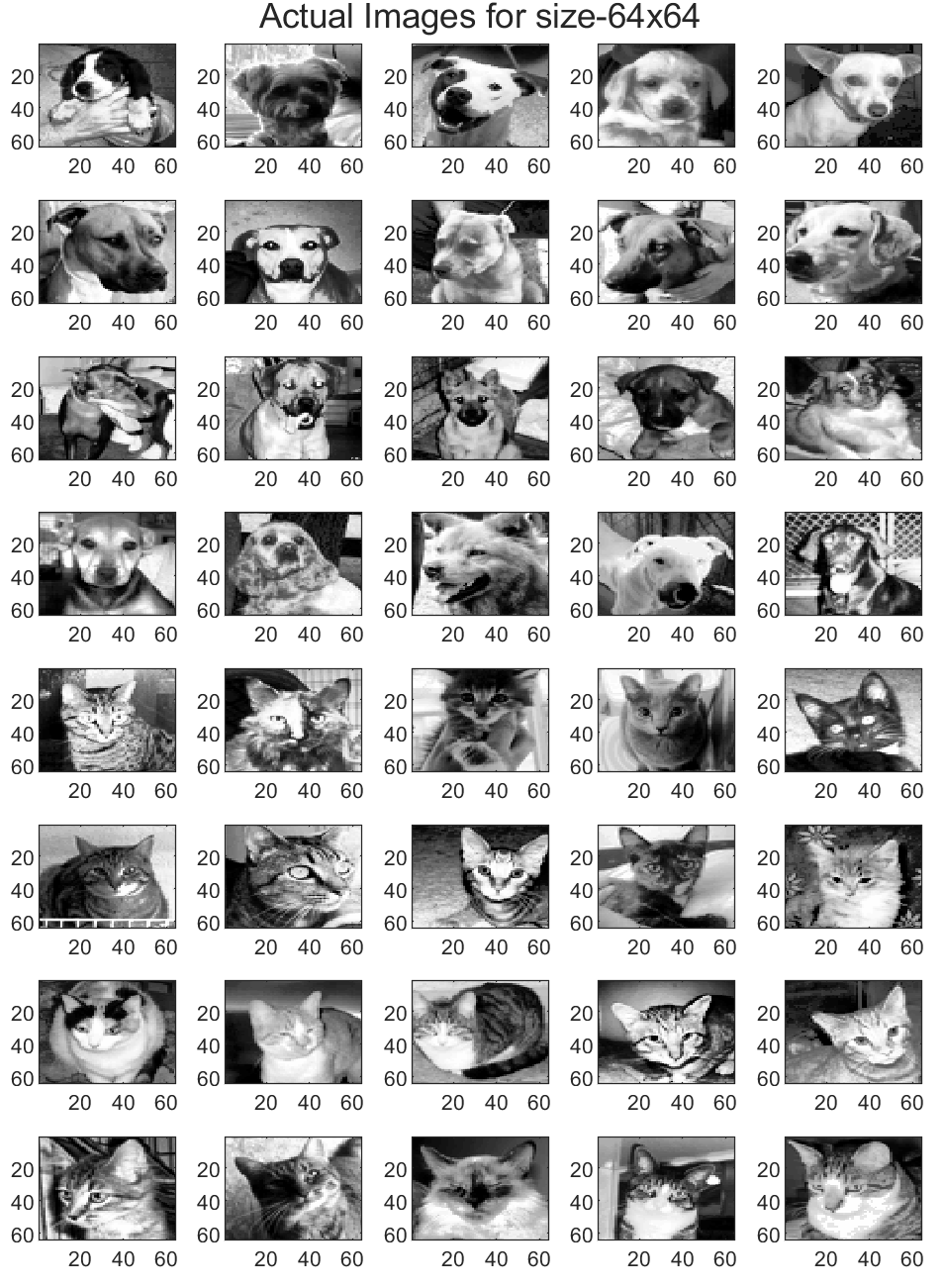}
    \caption{Dog and Cat Classification Images.}
    \label{20dogs20cats_64x64Actual_Images_cropped}
\end{figure}

Another similar classification problem uses images of cats and dogs as .jpg files in an image database.  Twenty of each are shown in Fig. \ref{20dogs20cats_64x64Actual_Images_cropped}.

As with the the letter classification application above, each image is transformed into a n-qubit ( $2^n$x$2^n$ ) density matrix using the quantum transform procedure above. Each density matrix then becomes the ``input'' to the Discriminator/classifier by assigning the initial state of the quantum Discriminator at $t_0$ to be that density matrix.  The Discriminator quantum dynamics propagate that initial state to a final state at $t_f$.  A measure of the correlation of that final state to the n-qubit Pauli state $\sigma_{zz}$ is made.  This becomes the output of the Discriminator/classifier.  This measure value is compared to a binary separator value to achieve the binary classification of ``diagonal left'' (below the separator) and ``diagonal right'' (above the separator).  A target output is defined to be above, or below, the separator, depending upon what is the correct classification.  The Hamiltonian parameters $K$, $\epsilon$, $\zeta$ and the Lindblad parameter $\Gamma$ are trained.  
\subsubsection{4 qubit 16x16 Image Results}
\begin{figure}
    \centering
    \includegraphics[width=1\linewidth]{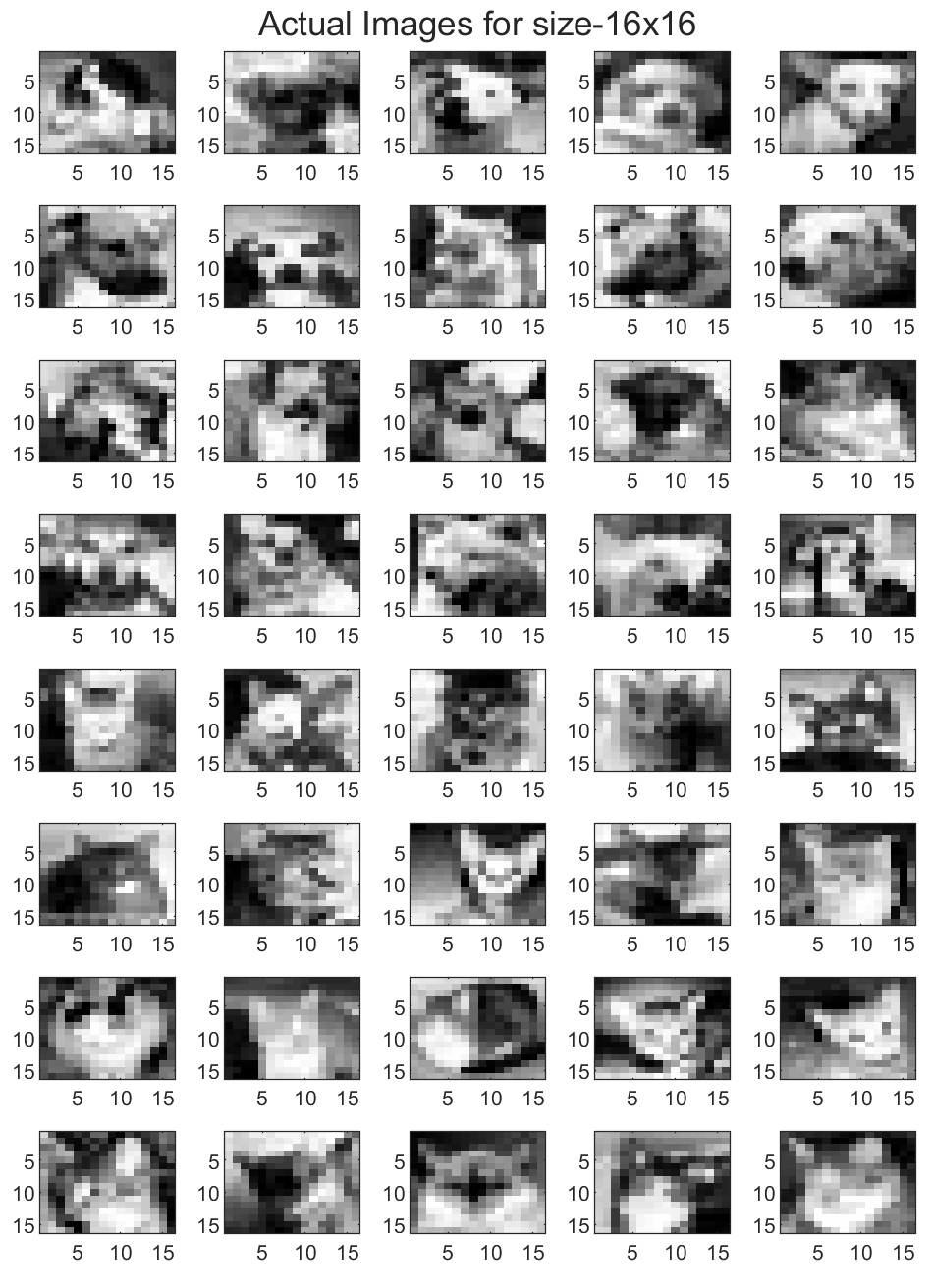}
    \caption{4 qubit 16x16 Pixel Dog and Cat Images}
    \label{20dogs20cats_16x16Actual_Images_cropped}
\end{figure}

\begin{figure}[htbp]
\centerline{\includegraphics[width=1\linewidth]{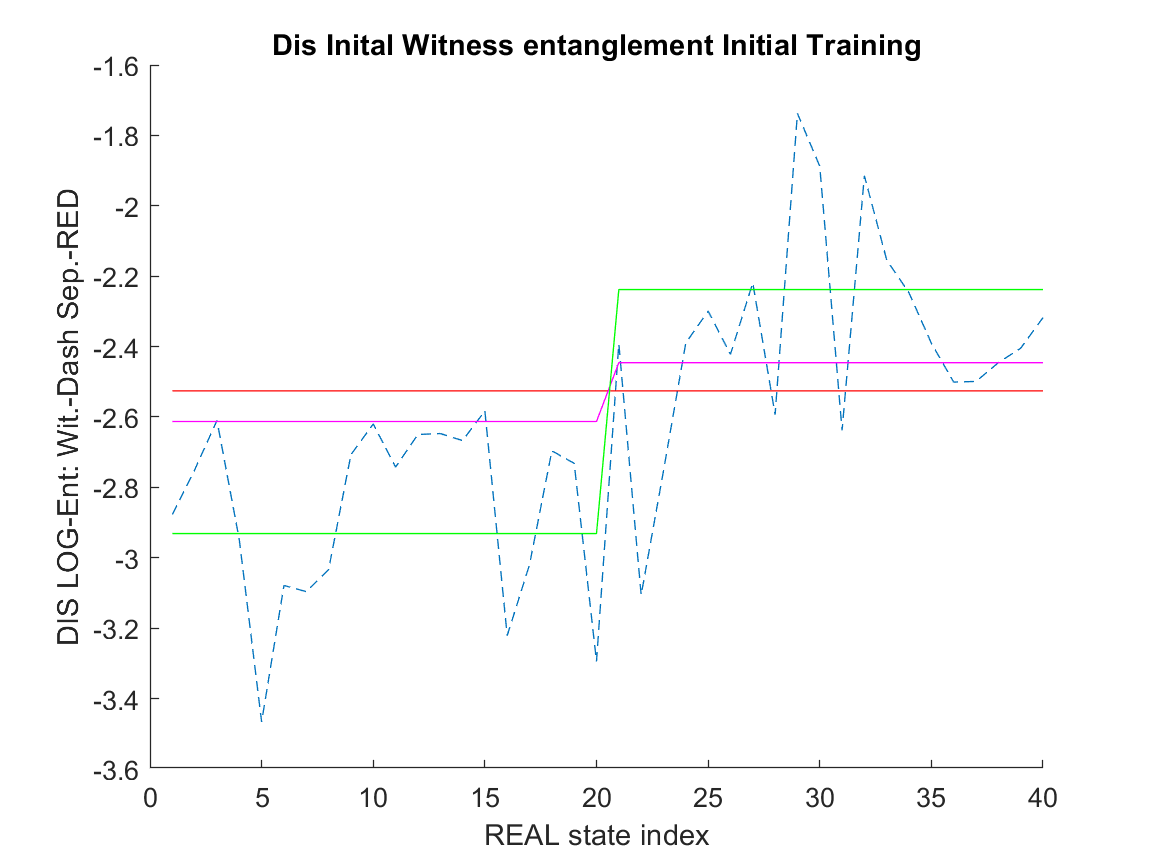}}
\caption{2 qubit Discriminator Training Classification. Red line is separator, Green line is Discriminator target measure values, Magenta line is 1/4 the distance to the target green line, Blue Dashed line is actual measure values.}
\label{20dogs20cats_4qubits_Training_Results}
\end{figure}

\begin{figure}[htbp]
\centerline{\includegraphics[width=1\linewidth]{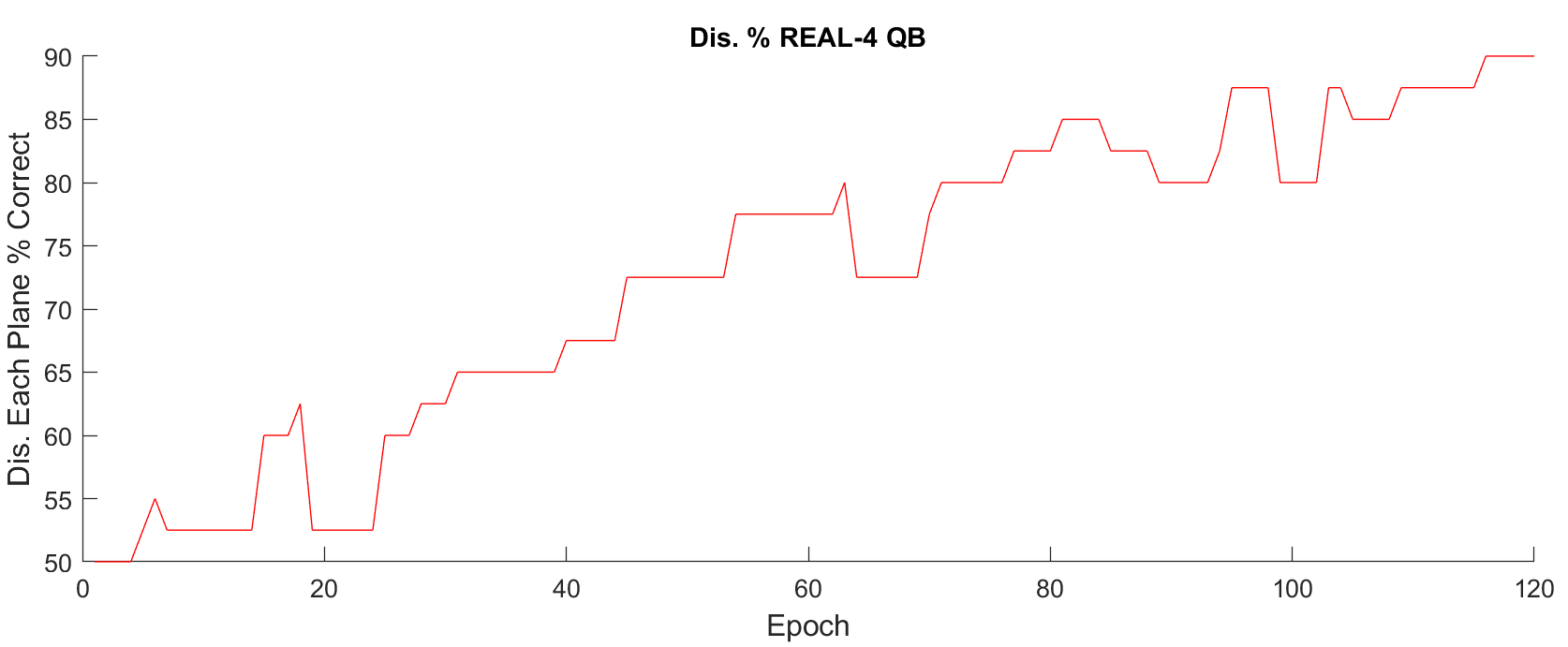}}
\caption{Discriminator Training Results Percent Classification -vs- Epoch.}
\label{20dogs20cats_4qubits_Percent_Classification}
\end{figure}

Down sampling the letter images for recognition with 4 qubits gives a very coarse 16x16 pixel resolution.  The down sampled images that are used, twenty for each of cats and dogs, are shown in Fig. \ref{20dogs20cats_16x16Actual_Images_cropped}
Following quantum image transformation, the discriminator is tasked with classifying them.  Training results are shown in Fig. \ref{20dogs20cats_4qubits_Training_Results}. The separator value is shown as the red line and the magenta line is 1/4 the distance from the separator to the target green line.The target values for the measured Discriminator output are shown as the green line  and the after training Discriminator output measure values are shown as the blue dotted line. During training, if an input is classified at or beyond the green target value the error used for training is discounted by a factor of 100 as this input is clearly correctly classified.  If an input is classified between the green target value and the magenta value, the error used for training is discounted by a factor of 5 as this input is  correctly, but not clearly classified. if an input is classified on the side of the magenta line nearest the separator, the error used for training is not discounted at all as this input is near enough to being NOT correctly classified. The $x$-axis is the image pair number.  Dogs are numbers 1-20 and cats are numbers 21-40. A plot of the RMS error during Discriminator training is shown in Fig. \ref{20dogs20cats_4qubits_Percent_Classification}.

\section{DISCUSSIONS AND CONCLUSIONS}

In this paper, we have applied our prior machine learning work to create and train a quantum GAN. We have demonstrated that it is a fully quantum network by showing that it can classify input quantum states on the basis of their state of entanglement. This alone is significant, as the determination of separability for systems larger than two qubits is an open NP-hard problem, and one of great interest to quantum computing. It is possible that the successful training of quantum networks to do problems like this may lead to greater analytical understanding, as well.

Next, we trained a binary image classifier via a novel quantum image transform that transforms an MxM image into an N qubit quantum density matrix where N = $\log_2(M)$ (a 1024x1024 image is represented by 10 qubits).  And third, we created a quantum analog to the classical stylenet GANs developed by Karras for image generation and classification. A quantum system is used as a Generator and a separate quantum system is used as a Discriminator.  The Generator Hamiltonian quantum parameters are augmented by quantum style parameters which play the role of the style parameters used by Karras. 
%Both are trained in a GAN MiniMax problem along with the Discriminator Hamiltonian quantum parameters to generate and discriminate “real” quantum product states from “fake” quantum product states generated by the quantum Generator.  This demo problem of detecting quantum product states is purely quantum mechanical and has no classical analog. It is an open NP-hard problem for quantum product states of more than 2 qubits.
With aforementioned encoding of image pixels into quantum states as density matrices, the method demonstrated here for this quantum problem is also applicable to GAN image generation and detection, the problem so successfully demonstrated by Karras’s classical GAN. 
%We apply our prior work where we developed a modified quantum Levenberg-Marquardt (LM) machine learning technique for quantum computing. 

An advantage of machine learning is that it can be used as a systematic method to non-algorithmically “program” quantum computers. Quantum machine learning enables us to perform computations without breaking down an algorithm into its gate “building blocks”, eliminating that difficult step and potentially reducing unnecessary complexity. In addition, the machine learning approach is robust to both noise and to decoherence, which is ideal for running on inherently noisy NISQ devices which are limited in the number of qubits available for error correction.

While this method works extremely well to train quantum systems in simulation, it requires knowledge of the quantum states,  the density matrix $\rho$ at each time $t$.  This makes this method not amenable to real time quantum hardware training, since measuring the quantum state at intermediate times collapses the quantum state and destroys the quantum mechanics computation \cite{origb19}.  In other words, our machine learning method, that we have called ``quantum backprop'' in early papers, can be run in simulation and the resulting approximate parameters executed on quantum hardware as we have done in \cite{origb20}, but training on the hardware itself cannot be accomplished.  Also, because the $H$ used in the above “off-line” machine learning does not exactly match the quantum hardware due to unknown dynamics and uncertainties in the physical system, the resulting calculations on the hardware will have some error.  The “off-line” parameters can, however, be a starting point for further machine learning refinement on the hardware as described in \cite{Steck_Behrman_Thompson}.  
%Mention work using "partial" state measurement for actor critic methods. [???]

A major contribution of this paper is the demonstration of the feasibility of true online training of a quantum system to do a quantum calculation. It is a well-known theorem that a very small set of gates (e.g., the set {H, T, S, CNOT}) is universal. This means that any N-qubit unitary operation can be approximated to an arbitrary precision by a sequence of gates from that set. But there are many calculations we might like to do, for which we do not know an optimal sequence to use, or even, perhaps, any sequence to use. And there are many questions we might want to answer for which we do not even have a unitary, that is, an algorithm. 
While this paper uses a Quantum GAN to process product state, which can naturally be represented as a quantum state, creation of a Quantum GAN (QGAN) brings this algorithm to the quantum space of machine learning, opening up the ability to generate and classify high resolution images with as few as 12 qubits.  In simulation, 10 qubits is still very challenging, yet, in hardware available in NISQ, this would be a significant potential improvement in processing speed and image representation memory.  It still remains to be shown that we can encode an image as a quantum state efficiently, yet the promise is there.

\section{ACKNOWLEDGMENT}
We all thank the entire research group for helpful discussions: Nam Nguyen, William Ingle, and Ricardo Rodriguez.

\end{document}